\documentclass[acmsmall]{acmart}

\acmJournal{TALG}
\acmVolume{0}
\acmNumber{0}
\acmArticle{0}
\acmYear{YYYY}
\acmMonth{0}

\newtheorem{problem}[theorem]{Problem}
\newtheorem{remark}[theorem]{Remark}

\usepackage{amsmath, amssymb}
\usepackage{bm}
\usepackage{algorithm}
\usepackage{algorithmicx,algpseudocode}
\usepackage{color}
\usepackage{tikz}
\usetikzlibrary{calc}
\usepackage{multirow}
\usepackage{cancel}
\usepackage{soul}

\usepackage{sidecap}
\sidecaptionvpos{figure}{c}

\algloopdefx{NFor}[1]{\textbf{for} #1 \textbf{do}}
\algloopdefx{NIf}[1]{\textbf{if} #1 \textbf{then}}

\newcommand{\meas}{{\mathbf \mu}}
\newcommand{\signal}{{\mathbf x}}

\newcommand{\poly}{\operatorname{poly}}

\newcommand{\mtx}[1]{\bm{#1}}
\newcommand{\supp}[1]{\operatorname{supp}(#1)}
\newcommand{\nerr}[1]{\left\| #1 \right\|}
\newcommand{\mb}{\ensuremath\mathbf}
\newcommand{\F}{\ensuremath\mathbb{F}}

\DeclareMathOperator*{\median}{median}

\title{For-all Sparse Recovery in Near-Optimal Time}
\author{ANNA C. GILBERT}
\affiliation{%
  \institution{Department of Mathematics. University of Michigan, Ann Arbor}
  \streetaddress{530 Church St}
  \city{Ann Arbor}
  \state{MI}
  \postcode{48109}
  \country{USA}}
\email{annacg@umich.edu}
\author{YI LI}
\affiliation{%
  \institution{Division of Mathematics, SPMS. Nanyang Technological University}
  \streetaddress{21 Nanyang Link}
  \postcode{637371}
  \country{Singapore}}
\email{yili@ntu.edu.sg}
\author{ELY PORAT}
\affiliation{
  \institution{Department of Computer Science. Bar-Ilan University}
  \city{Ramat-Gan}
  \postcode{5290002}
  \country{Israel}}
\email{porately@cs.biu.ac.il}
\author{MARTIN J. STRAUSS}
\affiliation{
  \institution{Department of Mathematics. University of Michigan, Ann Arbor}
  \streetaddress{530 Church St}
  \city{Ann Arbor}
  \state{MI}
  \postcode{48109}
  \country{USA}}
\email{martinjs@umich.edu}

\begin{abstract}
An \textit{approximate sparse recovery system} in $\ell_1$ norm consists of parameters $k$, $\epsilon$, $N$, an $m$-by-$N$ measurement $\Phi$, and a recovery algorithm, $\mathcal{R}$. Given a vector, $\mb{x}$, the system approximates $x$ by $\widehat{\mb{x}} = \mathcal{R}(\Phi\mb{x})$, which must satisfy $\|\widehat{\mb{x}}-\mb{x}\|_1 \leq (1+\epsilon)\|\mb{x}-\mb{x}_k\|_1$. We consider the ``for all'' model, in which a single matrix $\Phi$, possibly ``constructed'' non-explicitly using the probabilistic method, is used for all signals $\mb{x}$. The best existing sublinear algorithm by Porat and Strauss (SODA'12) uses $O(\epsilon^{-3} k\log(N/k))$ measurements and runs in time $O(k^{1-\alpha}N^\alpha)$ for any constant $\alpha > 0$. 

In this paper, we improve the number of measurements to $O(\epsilon^{-2} k \log(N/k))$, matching the best existing upper bound (attained by super-linear algorithms), and the runtime to
$O(k^{1+\beta}\poly(\log N,1/\epsilon))$, with a modest restriction that $k\leq N^{1-\alpha}$ and $\epsilon \leq (\log k/\log N)^{\gamma}$, for any constants $\alpha, \beta,\gamma > 0$. When $k\leq \log^c N$ for some $c>0$, the runtime is reduced to $O(k\poly(N,1/\epsilon))$. With no restrictions on $\epsilon$, we have an approximation recovery system with $m = O(k/\epsilon \log(N/k)((\log N/\log k)^\gamma + 1/\epsilon))$ measurements.

The overall architecture of this algorithm is similar to that of Porat and Strauss (SODA'12) in that we repeatedly use a weak recovery system (with varying parameters) to obtain a top level recovery algorithm. The weak recovery system consists of a two-layer hashing procedure (or with two unbalanced expanders, for a deterministic algorithm). The algorithmic innovation is a novel encoding procedure that is reminiscent of network coding and that reflects the structure of the hashing stages. The idea is to encode the signal position index $i$ by associating it with a unique message $\mb{m}_i$, which will be encoded to a longer message $\mb{m}_i'$ (in contrast to (Porat-Strauss, SODA'12) in which the encoding is simply the identity). Portions of the message $\mb{m}_i'$ correspond to repetitions of the hashing and we use a regular expander graph to encode the linkages among these portions. 

The decoding or recovery algorithm consists of recovering the portions of the longer messages $\mb{m}_i'$ and then decoding to the original messages $\mb{m}_i$, all the while ensuring that corruptions can be detected and/or corrected. The recovery algorithm is similar to list recovery introduced in (Indyk et al., SODA'10) and used in (Gilbert et al., ICALP'13). In our algorithm, the messages $\{\mb{m}_i\}$ are independent from the hashing, which enables us to obtain a better result.
\end{abstract}

\ccsdesc[500]{Theory of computation~Sketching and sampling}

\terms{Algorithms, Theory}
\keywords{Compressive sensing, list decoding, sparse recovery}

\begin{document}
\thanks{A preliminary version appeared in the Proceedings of ICALP 2014, \textit{LNCS} 8572, pp 538--550.\\
A. C. Gilbert was supported in part by DARPA/ONR N66001-08-1-2065. Y. Li was supported by NSF CCF 0743372 when he was at University of Michigan. M. J. Strauss was supported in part by  NSF CCF 0743372 and DARPA/ONR N66001-08-1-2065.
}
\maketitle

\section{Introduction}

Sparse signal recovery is a critical data-acquisition and processing problem that arises in many modern scientific and computational applications, including signal and image processing, machine learning, data networking, and medicine~\cite{DDT+08,LDP07}. It is a method for acquiring linear measurements or observations of a signal with a measurement matrix $\Phi$, and an algorithm, $\mathcal{D}$, for recovering the significant components of the original signal. We model this problem mathematically by assuming that we \emph{measure} a vector $\mb{x}$ and collect observation $\mb{y}=\Phi \mb{x}$, then we run a \emph{recovery algorithm} and produce an approximation $\widehat{\mb{x}}=\mathcal{D}(\Phi,\mb{y})$ to $\mb{x}$ with the guarantee that the approximation error $\|\widehat{\mb{x}}-\mb{x}\|$ is bounded above. 

More quantitatively, let us denote the length of the vector $\mb{x}$ by $N$, the sparsity (or compression) parameter $k$, and distortion parameter
$\epsilon$.  Let $\mb{x}_{[k]}$ denote the best $k$-term approximation to $\mb{x}$, the ``heavy hitters'' of $\mb{x}$, {\em i.e.}, $\mb{x}$ with all but the $k$ largest-magnitude terms zeroed out. There are many different ways to assess the error of the recovery algorithm and the quality of the measurement matrix, depending on the particular application. (See Table~\ref{table:previous} for an overview of all of problem variations.) In this paper, we address the $\ell_1/\ell_1$-forall problem\footnote{More generally, the expression $\ell_p/\ell_q$ means that we measure the approximation error $\|\widehat{\mb{x}} - \mb{x}\|_p$ with the $\ell_p$ norm and we compare it to the $\ell_q$ error of the best $k$-term approximation, $\|\mb{x}_{[k]}-\mb{x}\|_q$.} which is to
give a measurement matrix $\Phi$ and a recovery algorithm $\mathcal{D}$, such that, for any input vector $\mb{x}$, we have
\[
	\|\widehat{\mb{x}}-\mb{x}\|_1\le(1+\epsilon)\|\mb{x}_{[k]}-\mb{x}\|_1.
\]
The goal is to use the minimum number of measurements (rows of
$\Phi$), namely, $O(k\log(N/k)/\epsilon^2)$ and to keep the runtime of
$\mathcal{D}$ to polynomial in $k\log(N)/\epsilon$. Since the measurement matrix $\Phi$ is chosen independently of the input vector $\mb{x}$, it corresponds to \emph{non-adaptive} measurements.  (We do not know whether adaptivity would help in this setting.)

What makes this problem challenging is that we must simultaneously keep the number of measurements small, ensure the recovery algorithm is highly efficient, and achieve a good approximation for all input vectors. If we increase the number of measurements by factors of $\log N$, it is easy to optimize the run-time~\cite{BGIKS,Cheraghchi:2016:NOD:2884435.2884458}.
Similarly, if $\epsilon<1/N$, the desired bound allows at least $1/\epsilon > N$ measurements and the problem becomes trivial.
In many applications, all three quantities are important; i.e., in medical imaging applications, the measurements reflect the time a patient is observed, the recovery time drives the effectiveness of real-time imaging systems, and the recovery accuracy determines the diagnostic effectiveness of the imaging system.

\begin{table}
\caption{Summary of the best previous results and the
  result obtained in this paper.\label{table:previous}}{\footnotesize
  \begin{tabular}{|c|c|c|c|c|c|c|}
\hline
Paper           & A/E & Number of  & Column sparsity/ & Decode time  & 	Approx. error & Noise\\
 & &  Measurements & Update time & & & \\
\hline
\cite{CCFC02}      & E   & $k \log^{O(1)} N$      & $\log^{O(1)} N$       & $N \log^{O(1)} N$ & $\ell_2 \le C \ell_2$ & \\
\hline
\cite{CM06}
                & E   & $k \log^{O(1)} N$      & $\log^{O(1)} N$       & $k \log^{O(1)} N$ & $\ell_2 \le C \ell_2$ & \\
\hline
\cite{GLPS}   & E   & $\epsilon^{-1}k \log(N/k)$     & $\log^{O(1)} N$       & $\epsilon^{-1}k\log^{O(1)} N$  & $\ell_2 \le (1+\epsilon) \ell_2$ & Y \\
\hline
\cite{Donoho06,CRT06}
                & A   & $k\log( N/k)$     & $k\log(N/k)$     & LP           & $\ell_2 \le (C/\sqrt{k}) \ell_1$ & Y \\
\hline
\cite{GSTV07:HHS}
                & A   & $\epsilon^{-2}k\log^{O(1)} N$       & $\epsilon^{-2}k\log^{O(1)} N$      & $\epsilon^{-4}k^2\log^{O(1)} N$& 
$\ell_2\le (\epsilon/\sqrt{k})\ell_1$   & Y \\
\hline
\cite{GSTV06}   & A   & $k\log^{O(1)} N$       & $\log^{O(1)} N$       & $k\log^{O(1)} N$  & $\ell_1\le (C\log N)\ell_1$   & Y \\
\hline
\cite{IR08}     & A   & $\epsilon^{-2}k\log(N/k)$      & $\epsilon^{-1}\log(N/k)$      & $N\log(N/k)$ & $\ell_1\le(1+\epsilon)\ell_1$ & Y\\
\hline
\cite{PS12}		& A   & $\ell^{O(1)}\epsilon^{-3}k\log(N/k)$      & $\ell^{O(1)}\epsilon^{-3}\log(N/k)\log k$      & $\ell^{O(1)} \epsilon^{-3} k(N/k)^{1/\ell}$ & $\ell_1\le(1+\epsilon)\ell_1$ & Y\\
\hline
\hline
This paper
                & A   & $\epsilon^{-2}k\log N$
                                          & $\epsilon^{-1}\log N$
                                                             & $k^{1+\beta}(\epsilon^{-1}\log N)^{O(1)}$
                                                                            & $\ell_1\le(1+\epsilon)\ell_1$ & \\
\hline
\hline
Lower bound `A'                & A   & $\epsilon^{-2}k\log(N/k)$      & $\epsilon^{-1}\log(N/k)$      & $\epsilon^{-2}k\log(N/k)$ & $\ell_2\le(\epsilon/\sqrt{k})\ell_1$ & Y\\
\hline
\end{tabular}}
\footnotesize
Summary of the best previous results and the
  result obtained in this paper.  The measurement and time complexities are subject to $O$-notations, which are suppressed for clarity.  In~\cite{PS12}, $\ell$ is an arbitrary positive constant integer and the $O(1)$ in exponents are absolute constants; in the result of this paper, $\beta$ is an arbitrary positive constant, the $O(1)$ in the exponent in decode time takes the form of $c_1 + c_2\beta$ (where $c_1, c_2$ are absolute constants) and restrictions on $k$ and $\epsilon$ apply. ``LP'' denotes (at least) the time to solve a linear
  program of size at least $N$.  The column ``A/E'' indicates whether
  the algorithm works in the forall (A) model or the foreach (E)
  model.  The column ``noise'' indicates whether the algorithm
  tolerates noisy measurements, that is, the observation $\mb{y}=\Phi\mb{x}+\bm{\nu}$.  Measurement and decode time dependence on
  $\epsilon$, where applicable, is polynomial.  The lower bound on number of measurements in table above is, in fact, the best upper bound attained by super-linear algorithms.
\end{table}

\paragraph{Related work.} There has been considerable work on this problem in a variety of parameter settings and we summarize the results in Table~\ref{table:previous}. A number of parameter values are incommensurate: we can achieve better approximation guarantees (using the $\ell_2/\ell_2$ norm) but only in the for-each model and in the for-all signal model, we can achieve $\ell_2/\ell_1$ error guarantees. A somewhat harder problem than the one we address in this paper is the
mixed-norm (or $\ell_2/\ell_1$) for-all result.  In this setting, the goal is to give $\Phi$ and
$\mathcal{D}$, such that, for any $\mb{x}$, we have
\begin{equation}\label{eqn:mixed-norm}
\|\widehat{\mb{x}}-\mb{x}\|_2\le\frac{\epsilon}{\sqrt{k}}\|\mb{x}_{[k]}-\mb{x}\|_1.
\end{equation}
It is known that if $(\Phi,\mathcal{D})$ solves the $\ell_2/\ell_1$ problem it also solves the $\ell_1/\ell_1$ problem \cite{CDD09}.

In another direction, the $\ell_2/\ell_2$ for-each problem is to give
{\em distribution} $\mathcal{F}$ on $\Phi$,
and $\mathcal{D}$, such that, for any $\mb{x}$, if $\Phi\sim\mathcal{F}$,
we have
\[
	\Pr_{\Phi\sim\mathcal{F}}\left\{\|\widehat{\mb{x}}-\mb{x}\|_2   	    \le(1+\epsilon)\|\mb{x}_{[k]}-\mb{x}\|_2\right\} \geq 1 - O(1).
\]
The $\ell_2/\ell_2$ for-each problem with constant failure probability was solved in~\cite{GLPS}, where the authors gave an algorithm with
constant-factor-optimal runtime and number of measurements. The failure probability was recently improved to exponentially small in \cite{ICALP}, but the technique is not likely to give an $\ell_1/\ell_1$ for-all result without additional logarithmic factors in the number of measurements.

The first sublinear-time algorithm in the for-all setting (for the $\ell_1/\ell_1$ norm) was given in~\cite{PS12}, though that algorithm had a number of limitations.
\begin{itemize}
\item The runtime, while sublinear, was $\sqrt{kN}$ or, more
  generally, of the form $k^{1-\alpha}N^{\alpha}$ for any constant
  $\alpha>0$.  That algorithm does not achieve polynomial running time in
  $k\log(N)/\epsilon$.
\item The algorithm requires a precomputed table of size $Nk^{0.2}$.
\item The dependence on $\epsilon$ is $1/\epsilon^{3}$, far from optimal dependence of $1/\epsilon^2$.
\end{itemize}

\paragraph{Our results.} In this work, we rectify the above limitations, assuming the (modest) restriction that $\epsilon<\log k/\log N$. We also make the measurement dependence on $\epsilon$ optimal. The best lower bound for the $\ell_1/\ell_1$ for-all problem is $\Omega(k/\epsilon^2 + (k/\epsilon)\log(\epsilon N/k))$ \cite{NNW12}, which is also the best lower bound for the $\ell_2/\ell_1$ for-all problem. Our algorithm uses $O(k/\epsilon^2\log(N/k))$ measurements when $\epsilon < (\log k/\log N)^{\gamma}$, which is suboptimal only by a logarithmic factor. When $k\leq \log^c N$ for some $c>0$, the runtime is reduced to $O(k\poly(N,1/\epsilon))$.
\begin{theorem}[Main Theorem]
\label{thm:mainresult}
Let $\beta,\gamma > 0$. There is an \emph{approximate sparse recovery system} consisting of an $m \times N$ measurement matrix $\mtx{\Phi}$ and a decoding algorithm $\mathcal{D}$ that satisfy the following property: for any vector $\signal\in \mathbb{R}^n$, given
$\mtx{\Phi}\signal$, the system approximates $\signal$ by $\widehat \signal=\mathcal{D}(\mtx{\Phi}
\signal)$, which satisfies
\[
	\|\widehat \signal - \signal\|_1
	\le (1+\epsilon)\|\signal_{[k]} - \signal\|_1.
\]
Provided that $N=\Omega(\max\{k^2, k/\epsilon^2\})$, the matrix $\mtx{\Phi}$ has $m = O(k/\epsilon \log(N)((\log N/\log k)^\gamma + 1/\epsilon))$ rows and the decoding algorithm $\mathcal{D}$ runs in time $O(k^{1+\beta}\poly(\log N,1/\epsilon))$. When $\epsilon = O\bigl((\frac{\log k}{\log N})^\gamma\bigr)$, the number of rows is $m = O(k/\epsilon^2\log N)$. If, in addition, $k\leq \log^{O(1)} N$, the runtime can be reduced to $O(k\poly(\log N,1/\epsilon))$.
\end{theorem}

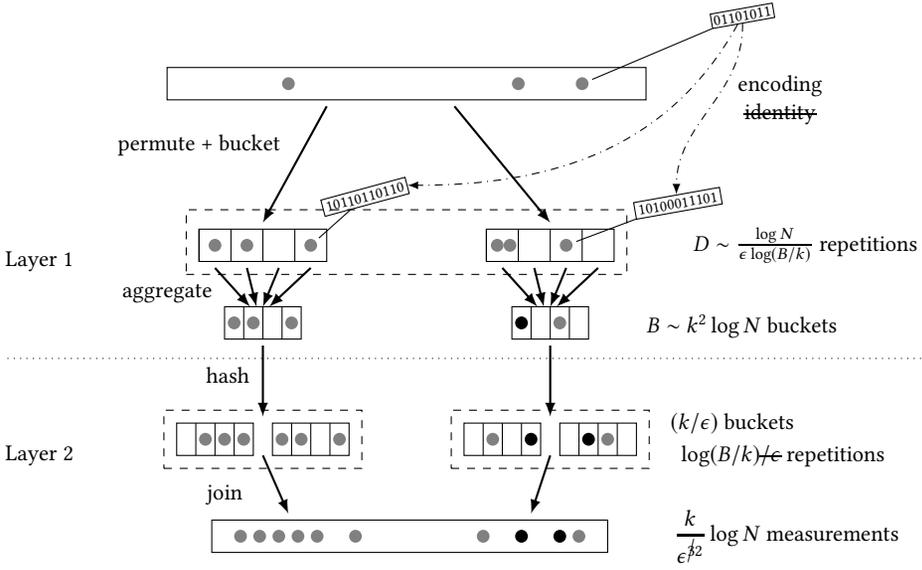
\begin{figure}
\begin{tikzpicture}[scale=0.85,every node/.style={transform shape}]
\coordinate (a) at (0,9.5);

\draw[thin] ($(a)+(-1.5,0)$) rectangle ($(a)+(6,0.5)$);
\foreach \i/\x in {1/0.4, 2/4, 3/5}{
\node[gray,fill,circle,inner sep=2pt,minimum size=5pt,outer sep=2pt] (a1\i) at ($(a)+(\x,0.25)$) {};
};
\node[rotate=10,rectangle,draw=black,inner sep=1pt] (a13label) at ($(a)+(7.51,1.3)$) {{\scriptsize 01101011}};
\draw (a13) -- (a13label);

\coordinate (b) at (-1,7);
\coordinate (xb) at (0.4,-1.2);
\draw[thin] (b) rectangle ($(b)+(2,0.5)$);
\draw[dashed] ($(b)+(-0.2,-0.2)$) rectangle ($(b)+(6.7,0.8)$);

\foreach \i in {0,1,2,3}{
	\draw ($(b)+(0.5*\i,0)$) -- ($(b)+(0.5*\i,0.5)$);
	\draw[thick,-latex] ($(b)+(0.25+0.5*\i,-0.1)$) -- ($(b)+(0.8+0.1*\i,-0.7)$);
}
\draw[arrows={-latex},thick] ($(a)+(1,-0.1)$) -- ($(b)+(1,0.6)$);
\foreach \i/\x in {1/0.25, 2/0.75, 3/1.75}{
\node[gray,fill,circle,inner sep=2pt,minimum size=5pt,outer sep=2pt] (a2\i) at ($(b)+(\x,0.25)$) {};
};
\node[rotate=15,rectangle,draw=black,inner sep=1.5pt] (a23label) at ($(b)+(2.6,1)$) {{\scriptsize 10110110110}};
\draw (a23) -- (a23label);
\draw ($(b)+(xb)$) rectangle ($(b)+(xb)+(1.2,0.5)$);
\node[gray,fill,circle,inner sep=2pt,minimum size=5pt,outer sep=2pt] at ($(b)+(xb)+(0.15,0.25)$) {};
\node[gray,fill,circle,inner sep=2pt,minimum size=5pt,outer sep=2pt] at ($(b)+(xb)+(0.45,0.25)$) {};
\node[gray,fill,circle,inner sep=2pt,minimum size=5pt,outer sep=2pt] at ($(b)+(xb)+(1.05,0.25)$) {};

\foreach \i in {1,2,3}{
	\draw ($(b)+(xb)+(0.3*\i,0)$) -- ($(b)+(xb)+(0.3*\i,0.5)$);
}
\node at ($(b)+(0,1.8)$) {permute + bucket};
\node at ($(b)+(-0.5,-0.5)$) {aggregate};

\coordinate (b) at (3.5,7);

\draw[thin] (b) rectangle ($(b)+(2,0.5)$);
\foreach \i in {0,1,2,3}{
	\draw ($(b)+(0.5*\i,0)$) -- ($(b)+(0.5*\i,0.5)$);
	\draw[thick,-latex] ($(b)+(0.25+0.5*\i,-0.1)$) -- ($(b)+(0.8+0.1*\i,-0.7)$);
}
\draw[arrows={-latex},thick] ($(a)+(3,-0.1)$) -- ($(b)+(1,0.6)$);
\foreach \i/\x in {1/0.17, 2/0.37, 3/1.25}{
\node[gray,fill,circle,inner sep=2pt,minimum size=5pt,outer sep=2pt] (a3\i) at ($(b)+(\x,0.25)$) {};
};
\node[rotate=10,rectangle,draw=black,inner sep=1.5pt] (a33label) at ($(b)+(3,0.9)$) {{\scriptsize 10100011101}};
\draw (a33) -- (a33label);
\draw ($(b)+(xb)$) rectangle ($(b)+(xb)+(1.2,0.5)$);
\node[black,fill,circle,inner sep=2pt,minimum size=5pt,outer sep=2pt] at ($(b)+(xb)+(0.15,0.25)$) {};
\node[gray,fill,circle,inner sep=2pt,minimum size=5pt,outer sep=2pt] at
($(b)+(xb)+(0.75,0.25)$) {};
\foreach \i in {1,2,3}{
	\draw ($(b)+(xb)+(0.3*\i,0)$) -- ($(b)+(xb)+(0.3*\i,0.5)$);
}
\node at ($(b)+(4,-1)$) {$B\sim k^2\log N$ buckets };
\node at ($(b)+(5,0.25)$) {$D \sim \frac{\log N}{\epsilon \log(B/k)}$ repetitions};

\draw[dashdotted,arrows=-{latex},out=-90,in=90] (a13label) to (a33label);
\draw[dashdotted,arrows=-{latex},out=240,in=0] (a13label) to (a23label);
\node[align=center] (encoding_text) at (8.1,9.5) {encoding\\ \st{identity}};

\draw[thin,dotted] (-4,5.5) -- (11,5.5);
\node at (-3.5,7) {Layer 1};
\node at (-3.5,4) {Layer 2};

\coordinate (b) at (-1.35,4);
\coordinate (xb) at (1.5,0);
\draw[thin] ($(b)$) rectangle ($(b)+(1.2,0.5)$);
\draw[thin] ($(b)+(xb)$) rectangle ($(b)+(xb)+(1.2,0.5)$);
\draw[thick,-latex] ($(b)+(1.35,1.7)$) -- ($(b)+(1.35,0.6)$);
\node at ($(b)+(0.8,1.25)$) {hash};
\foreach \i in {1,2,3}{
	\draw ($(b)+(0.3*\i,0)$) -- ($(b)+(0.3*\i,0.5)$);
	\draw ($(b)+(xb)+(0.3*\i,0)$) -- ($(b)+(xb)+(0.3*\i,0.5)$);
}

\node[gray,fill,circle,inner sep=2pt,minimum size=5pt,outer sep=2pt] at ($(b)+(0.75,0.25)$) {};
\node[gray,fill,circle,inner sep=2pt,minimum size=5pt,outer sep=2pt] at ($(b)+(0.45,0.25)$) {};
\node[gray,fill,circle,inner sep=2pt,minimum size=5pt,outer sep=2pt] at ($(b)+(1.05,0.25)$) {};
\node[gray,fill,circle,inner sep=2pt,minimum size=5pt,outer sep=2pt] at ($(b)+(xb)+(0.15,0.25)$) {};
\node[gray,fill,circle,inner sep=2pt,minimum size=5pt,outer sep=2pt] at ($(b)+(xb)+(0.45,0.25)$) {};
\node[gray,fill,circle,inner sep=2pt,minimum size=5pt,outer sep=2pt] at ($(b)+(xb)+(1.05,0.25)$) {};
\draw[dashed] ($(b)+(-0.2,-0.2)$) rectangle ($(b)+(2.9,0.7)$);

\coordinate (b) at (3.15,4);
\coordinate (xb) at (1.5,0);
\draw[thin] ($(b)$) rectangle ($(b)+(1.2,0.5)$);
\draw[thin] ($(b)+(xb)$) rectangle ($(b)+(xb)+(1.2,0.5)$);
\draw[thick,-latex] ($(b)+(1.35,1.7)$) -- ($(b)+(1.35,0.6)$);
\foreach \i in {1,2,3}{
	\draw ($(b)+(0.3*\i,0)$) -- ($(b)+(0.3*\i,0.5)$);
	\draw ($(b)+(xb)+(0.3*\i,0)$) -- ($(b)+(xb)+(0.3*\i,0.5)$);
}
\node[black,fill,circle,inner sep=2pt,minimum size=5pt,outer sep=2pt] at ($(b)+(1.05,0.25)$) {};
\node[gray,fill,circle,inner sep=2pt,minimum size=5pt,outer sep=2pt] at
($(b)+(0.45,0.25)$) {};
\node[black,fill,circle,inner sep=2pt,minimum size=5pt,outer sep=2pt] at ($(b)+(xb)+(0.45,0.25)$) {};
\node[gray,fill,circle,inner sep=2pt,minimum size=5pt,outer sep=2pt] at
($(b)+(xb)+(0.75,0.25)$) {};
\draw[dashed] ($(b)+(-0.2,-0.2)$) rectangle ($(b)+(2.9,0.7)$);
\node at ($(b)+(4.2,0.5)$) {$(k/\epsilon)$ buckets };
\node at ($(b)+(5,0)$) {$\log(B/k)\text{\st{$/\epsilon$}}$ repetitions};

\coordinate (c) at (-0.8,2.5);
\coordinate (d) at (3,2.5);
\draw (c) rectangle ($(d)+(0.3*8,0.5)$);
\draw[thick,-latex] ($(c)+(0.8,1.5)$) -- ($(c)+(0.3*4,0.6)$);
\draw[thick,-latex] ($(d)+(1.5,1.5)$) -- ($(d)+(0.3*4,0.6)$);
\node at ($(c) + (0.2,0.9)$) {join};
\node[gray,fill,circle,inner sep=2pt,minimum size=5pt,outer sep=2pt] at ($(c)+(0.75,0.25)$) {};
\node[gray,fill,circle,inner sep=2pt,minimum size=5pt,outer sep=2pt] at ($(c)+(0.45,0.25)$) {};
\node[gray,fill,circle,inner sep=2pt,minimum size=5pt,outer sep=2pt] at ($(c)+(1.05,0.25)$) {};
\node[gray,fill,circle,inner sep=2pt,minimum size=5pt,outer sep=2pt] at ($(c)+(0.3*4+0.15,0.25)$) {};
\node[gray,fill,circle,inner sep=2pt,minimum size=5pt,outer sep=2pt] at ($(c)+(0.3*4+0.45,0.25)$) {};
\node[gray,fill,circle,inner sep=2pt,minimum size=5pt,outer sep=2pt] at ($(c)+(0.3*4+1.05,0.25)$) {};
\node[black,fill,circle,inner sep=2pt,minimum size=5pt,outer sep=2pt] at ($(d)+(1.05,0.25)$) {};
\node[gray,fill,circle,inner sep=2pt,minimum size=5pt,outer sep=2pt] at
($(d)+(0.45,0.25)$) {};
\node[black,fill,circle,inner sep=2pt,minimum size=5pt,outer sep=2pt] at ($(d)+(0.3*4+0.45,0.25)$) {};
\node[gray,fill,circle,inner sep=2pt,minimum size=5pt,outer sep=2pt] at
($(d)+(0.3*4+0.75,0.25)$) {};

\node at ($(d)+(5.2,0.25)$) {$\displaystyle \frac{k}{\epsilon^{\cancel{3}2}}\log N$ measurements};
\end{tikzpicture}
\caption{Algorithm to generate the measurements. Darker spots indicate a bigger value of the bucket/measurement. Strikethroughs are used to show where our approach or our object sizes differ from \cite{PS12}.}
\label{fig:measure}
\end{figure}

\begin{figure}
\begin{tikzpicture}[scale=0.85,every node/.style={transform shape}]
\coordinate (a) at (0,7.5);

\draw[thin] ($(a)+(-1.5,0)$) rectangle ($(a)+(6,0.5)$);
\foreach \i/\x in {1/0.4, 2/4, 3/5}{
\node[gray,fill,circle,inner sep=2pt,minimum size=5pt,outer sep=2pt] (a1\i) at ($(a)+(\x,0.25)$) {};
};
\node[rotate=10,rectangle,draw=black,inner sep=1pt] (a13label) at ($(a)+(7.51,1.3)$) {{\scriptsize 01101011}};
\draw (a13) -- (a13label);

\coordinate (b) at (-1,7);
\coordinate (xb) at (0.4,-1.2);
\node[rotate=15,rectangle,draw=black,inner sep=1.5pt] (a23label) at ($(b)+(xb)+(2.6,1)$) {{\scriptsize 10110110110}};
\draw ($(b)+(xb)$) rectangle ($(b)+(xb)+(1.2,0.5)$);
\node[gray,fill,circle,inner sep=2pt,minimum size=5pt,outer sep=2pt] at ($(b)+(xb)+(0.15,0.25)$) {};
\node[gray,fill,circle,inner sep=2pt,minimum size=5pt,outer sep=2pt] at ($(b)+(xb)+(0.45,0.25)$) {};
\node[gray,fill,circle,inner sep=2pt,minimum size=5pt,outer sep=2pt] (a23) at  ($(b)+(xb)+(1.05,0.25)$) {};
\draw (a23) -- (a23label);

\foreach \i in {1,2,3}{
	\draw ($(b)+(xb)+(0.3*\i,0)$) -- ($(b)+(xb)+(0.3*\i,0.5)$);
}
\draw[thick,-latex] ($(b)+(xb)+(0.3*2,0.6)$)--($(b)+(xb)+(0.3*2+0.5,1.6)$);

\coordinate (b) at (3.5,7);
\draw[thick,-latex] ($(b)+(xb)+(0.3*2,0.6)$)--($(b)+(xb)+(0.3*2-0.5,1.6)$);
\node[rotate=10,rectangle,draw=black,inner sep=1.5pt] (a33label) at ($(b)+(xb)+(3,0.9)$) {{\scriptsize 10100011101}};
\draw ($(b)+(xb)$) rectangle ($(b)+(xb)+(1.2,0.5)$);
\node[black,fill,circle,inner sep=2pt,minimum size=5pt,outer sep=2pt] at ($(b)+(xb)+(0.15,0.25)$) {};
\node[gray,fill,circle,inner sep=2pt,minimum size=5pt,outer sep=2pt] (a33) at
($(b)+(xb)+(0.75,0.25)$) {};
\draw (a33) -- (a33label);

\foreach \i in {1,2,3}{
	\draw ($(b)+(xb)+(0.3*\i,0)$) -- ($(b)+(xb)+(0.3*\i,0.5)$);
}

\draw[thin,dotted] (-4,5.5) -- (11,5.5);
\node at (-3.5,7) {Layer 1};
\node at (-3.5,4) {Layer 2};

\coordinate (b) at (-1.35,4);
\coordinate (xb) at (1.5,0);
\draw[thin] ($(b)$) rectangle ($(b)+(1.2,0.5)$);
\draw[thin] ($(b)+(xb)$) rectangle ($(b)+(xb)+(1.2,0.5)$);
\draw[thick,latex-] ($(b)+(1.35,1.7)$) -- ($(b)+(1.35,0.6)$);
\foreach \i in {1,2,3}{
	\draw ($(b)+(0.3*\i,0)$) -- ($(b)+(0.3*\i,0.5)$);
	\draw ($(b)+(xb)+(0.3*\i,0)$) -- ($(b)+(xb)+(0.3*\i,0.5)$);
}

\node[gray,fill,circle,inner sep=2pt,minimum size=5pt,outer sep=2pt] at ($(b)+(0.75,0.25)$) {};
\node[gray,fill,circle,inner sep=2pt,minimum size=5pt,outer sep=2pt] at ($(b)+(0.45,0.25)$) {};
\node[gray,fill,circle,inner sep=2pt,minimum size=5pt,outer sep=2pt] at ($(b)+(1.05,0.25)$) {};
\node[gray,fill,circle,inner sep=2pt,minimum size=5pt,outer sep=2pt] at ($(b)+(xb)+(0.15,0.25)$) {};
\node[gray,fill,circle,inner sep=2pt,minimum size=5pt,outer sep=2pt] at ($(b)+(xb)+(0.45,0.25)$) {};
\node[gray,fill,circle,inner sep=2pt,minimum size=5pt,outer sep=2pt] at ($(b)+(xb)+(1.05,0.25)$) {};
\draw[dashed] ($(b)+(-0.2,-0.2)$) rectangle ($(b)+(2.9,0.7)$);

\coordinate (b) at (3.15,4);
\coordinate (xb) at (1.5,0);
\draw[thin] ($(b)$) rectangle ($(b)+(1.2,0.5)$);
\draw[thin] ($(b)+(xb)$) rectangle ($(b)+(xb)+(1.2,0.5)$);
\draw[thick,latex-] ($(b)+(1.35,1.7)$) -- ($(b)+(1.35,0.6)$);
\foreach \i in {1,2,3}{
	\draw ($(b)+(0.3*\i,0)$) -- ($(b)+(0.3*\i,0.5)$);
	\draw ($(b)+(xb)+(0.3*\i,0)$) -- ($(b)+(xb)+(0.3*\i,0.5)$);
}
\node[black,fill,circle,inner sep=2pt,minimum size=5pt,outer sep=2pt] at ($(b)+(1.05,0.25)$) {};
\node[gray,fill,circle,inner sep=2pt,minimum size=5pt,outer sep=2pt] at
($(b)+(0.45,0.25)$) {};
\node[black,fill,circle,inner sep=2pt,minimum size=5pt,outer sep=2pt] at ($(b)+(xb)+(0.45,0.25)$) {};
\node[gray,fill,circle,inner sep=2pt,minimum size=5pt,outer sep=2pt] at
($(b)+(xb)+(0.75,0.25)$) {};
\draw[dashed] ($(b)+(-0.2,-0.2)$) rectangle ($(b)+(2.9,0.7)$);

\draw[dashdotted,arrows={latex}-,out=-90,in=90] (a13label) to (a33label);
\draw[dashdotted,arrows={latex}-,out=240,in=0] (a13label) to (a23label);
\node[align=center] (encoding_text) at (8.1,7.5) {decoding\\ \st{look up table}};

\coordinate (c) at (-0.8,2.5);
\coordinate (d) at (3,2.5);
\draw (c) rectangle ($(d)+(0.3*8,0.5)$);
\draw[thick,latex-] ($(c)+(0.8,1.5)$) -- ($(c)+(0.3*4,0.6)$);
\draw[thick,latex-] ($(d)+(1.5,1.5)$) -- ($(d)+(0.3*4,0.6)$);
\node at ($(c) + (0.2,0.9)$) {split};
\node[gray,fill,circle,inner sep=2pt,minimum size=5pt,outer sep=2pt] at ($(c)+(0.75,0.25)$) {};
\node[gray,fill,circle,inner sep=2pt,minimum size=5pt,outer sep=2pt] at ($(c)+(0.45,0.25)$) {};
\node[gray,fill,circle,inner sep=2pt,minimum size=5pt,outer sep=2pt] at ($(c)+(1.05,0.25)$) {};
\node[gray,fill,circle,inner sep=2pt,minimum size=5pt,outer sep=2pt] at ($(c)+(0.3*4+0.15,0.25)$) {};
\node[gray,fill,circle,inner sep=2pt,minimum size=5pt,outer sep=2pt] at ($(c)+(0.3*4+0.45,0.25)$) {};
\node[gray,fill,circle,inner sep=2pt,minimum size=5pt,outer sep=2pt] at ($(c)+(0.3*4+1.05,0.25)$) {};
\node[black,fill,circle,inner sep=2pt,minimum size=5pt,outer sep=2pt] at ($(d)+(1.05,0.25)$) {};
\node[gray,fill,circle,inner sep=2pt,minimum size=5pt,outer sep=2pt] at
($(d)+(0.45,0.25)$) {};
\node[black,fill,circle,inner sep=2pt,minimum size=5pt,outer sep=2pt] at ($(d)+(0.3*4+0.45,0.25)$) {};
\node[gray,fill,circle,inner sep=2pt,minimum size=5pt,outer sep=2pt] at
($(d)+(0.3*4+0.75,0.25)$) {};

\end{tikzpicture}
\caption{Algorithm to recover from the measurements}
\label{fig:recover}
\end{figure}
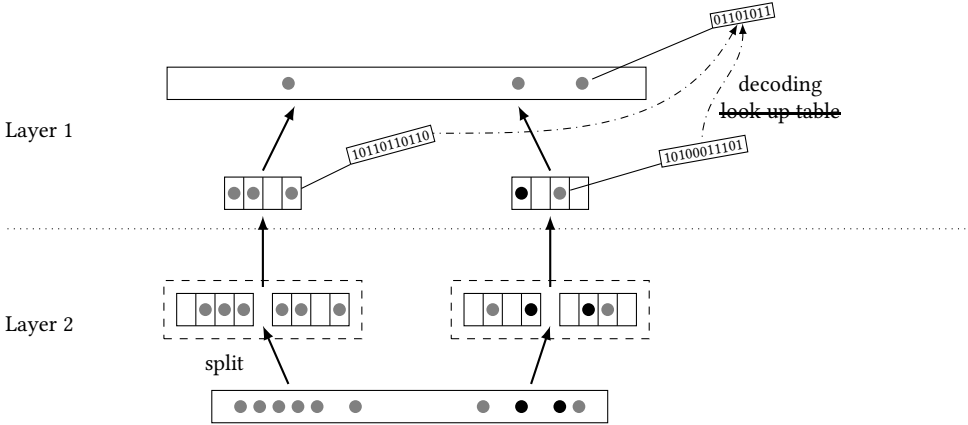

\begin{figure}
\begin{minipage}{0.48\textwidth}
\begin{tikzpicture}[scale=0.8,every node/.style={transform shape}]
\node[rectangle,draw=black] (original) at (3,5) {01101011};
\node[rectangle,draw=black] (PV) at (3,3.5) {1011 0110 1110};
\node[rectangle,draw=black] (chunk1) at (0.6,1.5) {1011\textbf{01}};
\node[rectangle,draw=black] (chunk2) at (3,1.5) {0110\textbf{11}};
\node[rectangle,draw=black] (chunk3) at (5.4,1.5) {1110\textbf{10}};
\node[rectangle,draw=black] (RS1) at (0.6,0) {011011011};
\node[rectangle,draw=black] (RS2) at (3,0) {110110010};
\node[rectangle,draw=black] (RS3) at (5.4,0) {0011010011};
\draw[-latex,thick] (original) -- (PV);
\foreach \i in {1,...,3}{
\draw[-latex,thick] (PV) -- (chunk\i);
\draw[-latex,thick] (chunk\i) -- (RS\i);
};
\node at (5.2,4.25) {PV code (outer encoding)};
\node[align=center] at (7.6,3.5) {$\epsilon^{-1}(\log N)/\log(B/k)$ blocks\\ each block has $\log(B/k)$ bits};
\node at (6.1,2.5) {$d$-regular expander};
\node[align=center] at (6.8,0.75) {Reed-Solomon\\ (inner encoding)};
\draw[<->,thick] (2,5.5)--(4,5.5);
\node[fill=white,rectangle,inner sep=1pt] at (3,5.5) {$\log N$};
\draw[<->,thick] (-0.4,-0.6)--(6.55,-0.6);
\node[fill=white,rectangle,inner sep=1pt] at (3,-0.6) {$\frac{C}{\epsilon}\log N$};

\draw[densely dashed] (3.9,3.5) ellipse (0.43 and 0.24);
\draw[densely dashed,out=0,in=180,->] (PV) to (5.2,3.5);
\end{tikzpicture}
\caption{Encoding scheme.  The Parvaresh-Vardy code automatically has a block structure. Suppose that there are $D$ blocks. Choose a $d$-regular expander on $D$ vertices as desired. For the $i$-th block of the PV code, append to it the information of the neighbours of the $i$-th vertex in the expander. Then apply Reed-Solomon to each appended message block. Note that the codes are non-binary.
}
\label{fig:encoding}
\end{minipage}
\hspace{0.02\linewidth}
\begin{minipage}{0.48\linewidth}
\begin{tikzpicture}[scale=0.8,every node/.style={transform shape}]
\node[rectangle,draw=black] (original) at (3,5) {01101011};
\node[rectangle,draw=black] (PV) at (3,3.5) {1011 0110 \xcancel{0010}};
\node[rectangle,draw=black] (chunk1) at (0.6,1.5) {1011\textbf{01}};
\node[rectangle,draw=black] (chunk2) at (3,1.5) {0110\textbf{11}};
\node[rectangle,draw=black] (chunk3) at (5.4,1.5) {\xcancel{0010\textbf{00}}};
\node[rectangle,draw=black] (RS1) at (0.6,0) {*1*011*11};
\node[rectangle,draw=black] (RS2) at (3,0) {1101***10};
\node[rectangle,draw=black] (RS3) at (5.4,0) {0**1*1**11};
\draw[latex-,thick] (original) -- (PV);
\foreach \i in {1,...,3}{
\draw[-latex,thick] (chunk\i) -- (PV);
\draw[-latex,thick] (RS\i) -- (chunk\i);
};
\node at (5.4,4.25) {PV decoding (outer decoding)};
\node[align=center] at (7.1,3.5) {obtaining a set of message blocks\\ of which a small fraction is good};
\node at (6.1,2.5) {cluster decoding};
\node[align=center] at (7.5,0.75) {Reed-Solomon decoding\\ (inner decoding)};

\end{tikzpicture}
\caption{Decoding scheme. The asterisks in the bottom layer indicate corrupted measurements (owing to collision or noise). The Reed-Solomon decoding either recovers the message block (with linking information) or produces a wrong one (crossed out) that is useless in recovering the original message. Then the clustering procedure finds a set of message blocks, of which a small fraction is good. This is sufficient for the Parvaresh-Vardy decoding to succeed.
}
\label{fig:decoding}
\end{minipage}
\end{figure}

\paragraph{Overview of Techniques.} Our overall approach builds on~\cite{PS12} and~\cite{ICALP} with several critical innovations. In Figure~\ref{fig:measure} is a framework which captures both the algorithm in \cite{PS12} and the algorithm in this paper. 

First, we describe the encoding procedure at a high level. Initially each $i\in [N]$ is associated with a unique message $\mb{m}_i$, which is encoded to a longer message $\mb{m}_i'$. In \cite{PS12} this encoding is trivial, namely, $\mb{m}_i' = \mb{m}_i$; while in our work it is a more complicated procedure (see Figure~\ref{fig:encoding}). The first hash assigns one of $B$ buckets to each $i\in[N]$, while maintaining the original index $i$; the {\em aggregation} step sums each bucket. There are
$\frac{\log N}{\epsilon\log(B/k)}$ repetitions. The index $i$ in each repetition is now associated with a block of $\mb{m}_i'$. In \cite{PS12}, the aggregated buckets are hashed into $(k/\epsilon)$ buckets and there
are $\log(B/k)/\epsilon$ repetitions.  Thus, altogether, there are
$O(\epsilon^{-3}k\log N)$ measurements (recall that $\log N = \Theta(\log(N/k))$ when $k = O(\sqrt{N})$). In our work, there are only $\log(B/k)$ repetitions, saving a factor of $1/\epsilon$, so the total number of measurements is $O(\epsilon^{-2}k\log N)$.

The \emph{identification} portion of the recovery algorithm is shown in Figure~\ref{fig:recover}. To recover the identity of heavy hitters, the algorithm reads off the measurements and recovers the message block associated with each bucket. This message block is supposed to be associated with the heavy hitter in the bucket. Then, all $B$ buckets are examined exhaustively. The pre-image of each heavy bucket under the first hash is determined, in \cite{PS12}, from a look-up table and searched exhaustively. In our work, this is done by the decoding procedure illustrated in Figure~\ref{fig:decoding}. We encode the ``linking information'' into the message blocks so that we can collect across the repetitions enough heavy buckets which contain the same heavy hitter $i$ (whose actual value is unknown at this stage of the algorithm). Thus, we obtain a (small) fraction of $\mb{m}_i'$, which is sufficient for the Parvaresh-Vardy decoding algorithm to produce the exact $\mb{m}_i$, from which we recover the value of $i$ immediately. 

The \emph{estimation} portion of the recovery algorithm estimates the coefficient at each of those candidate positions by reading the aggregated bucket value of the corresponding heavy buckets at the first hash level.

Putting these pieces together, we have a {\em weak recovery system}, which identifies all but $k/2$ of the heavy hitters.  We then repeat with smaller (easier) sparsity parameter $k/2<k$ and smaller (harder) distortion parameter
$(3/4)\epsilon<\epsilon$, resulting in a number of measurements whose leading
term is $(k/2)(4/3\epsilon)^2=(8/9)k/\epsilon^2<k/\epsilon^2$.
Summing the geometric progression gives the result we need. Finally, we note that our algorithm works (deterministically) with any unbalanced expander having the appropriate properties.

\paragraph{Encoding and Decoding details.} See Figure~\ref{fig:encoding} and Figure~\ref{fig:decoding} for a detailed illustration of these steps. For each message $\mb{m}$, the Parvaresh-Vardy code\footnote{There is no particular reason why we have chosen the Parvaresh-Vardy code; it can be replaced with other codes with similar or better performance, e.g., folded Reed-Solomon code.} encodes it into a longer message $\mb{m}'$, which automatically exhibits a block structure, so that if a few number of the blocks are correct, the original $\mb{m}$ will be recovered. Suppose there are $D$ blocks. Now, choose a $d$-regular expander graph $G$ ($d$ is a constant) on $D$ nodes such that after removing $O(D)$ nodes from $G$, the remaining graph still contains an expander of size $\Omega(D)$. For the $i$-th block of $\mb{m}'$, append to it the information of the neighbours of the $i$-th vertex in $G$. Then we apply Reed-Solomon to protect the appended blocks.

To decode, we first recover the appended message blocks. The two-layer hash guarantees that for the same heavy hitter, at most $O(D)$ of them will be wrong and the remaining ones are all correct. Now, consider a breadth-first search from a correct message block (whose ``linking information'' is therefore correct). By the special property of the expander graph $G$, we shall be able to visit all nodes (i.e., all corresponding message blocks) of a smaller expander graph of size $\Omega(D)$ in $\log D$ steps. This small fraction of good message blocks of $\mb{m}'$ will enable the P-V code to recover the original message $\mb{m}$ successfully. Recall that $d$ is a constant, the total number of vertices visited is $O(d^{\log D}) = O(\poly(D)) = O(\poly(\log N))$ for appropriate $D$. This enables a sublinear recovery time.

\paragraph{Our contributions.} 
\begin{itemize}
\item We give an algorithm for sparse recovery in the for-all setting,
  under a modest restriction on the distortion factor $\epsilon$, having the
  number of measurements that matches the best upper bound, attained by 
  super-linear algorithms; e.g., \cite{IR08}, and optimal in runtime up to a 
  power.
\item Our work is not the first to consider list recovery. Indyk et al.\ introduces the idea in the context of combinatorial group testing \cite{INR10}.   List recovery is also used in~\cite{DBLP:journals/dam/Cheraghchi13}.
The list recovery used in \cite{ICALP}, however, would affect the hashing and the hashing was thus required to be sufficiently random. In our algorithm, the messages $\{\mb{m}_i\}$ are independent of the hashing, which enables us to obtain a better result. 
\item Finally, our encoding/decoding techniques are reminiscent of network coding and may have other contexts for soft-decoding or network coding.
\end{itemize}

\paragraph{Paper Organization.} In Section~\ref{sec:prelim} we review some properties of expanders.
In Section~\ref{sec:weak}, we show that provided with good identification results, unbalanced expanders with appropriate properties will give a weak system. 
Our construction of weak system culminates in Section~\ref{sec:backpointers}, where we shall show how to achieve good identification via message encoding and decoding. Then we build the overall algorithm on the weak system in Section~\ref{sec:toplevel}. Finally we close with a short discussion and open problems in Section~\ref{sec:closing}.

\section{Preliminaries}\label{sec:prelim}

Our main algorithm will be built on regular graph expanders and unbalanced bipartite expanders (or rather, the adjacency matrices of such graphs). In an abuse of terminology, we will also use two different types of hashing schemes which can be implemented as (random) unbalanced bipartite expanders. In some contexts, it is more natural to describe and to analyze the structures as hashing schemes and in others, it is more natural to use the properties of expanders. In this section we review some properties of expanders and define precisely our hashing schemes. We also show that, up to an appropriate interpretation of the parameters, the two combinatorial structures are equivalent.

\subsection{Expander graphs}
Let $n,m,d,\ell$ be positive integers and $\epsilon,\kappa$ be positive reals. The following two definitions are adapted from \cite{GUV09}.

\begin{definition}[expander]
An $(n,\ell,\kappa)$-expander is a graph $G(V,E)$, where $|V|=n$, such that for any set $S\subseteq V$ with $|S|\leq \ell$ it holds that $|\Gamma(S)|\geq \kappa|S|$.
\end{definition}

\begin{definition}[bipartite expander]
An $(n,m,d,\ell,\epsilon)$-bipartite expander is a $d$-left-regular bipartite graph $G(L\cup R, E)$ where $|L| = n$ and $|R| = m$ such that for any $S\subseteq L$ with $|S|\leq \ell$ it holds that $|\Gamma(S)|\geq (1-\epsilon)d|S|$, where $\Gamma(S)$ is the neighbour of $S$ (in $R$).
\end{definition}

When $n$ and $m$ are clear from the context, we abbreviate the expander as $(\ell,d,\epsilon)$-bipartite expander.

Consider the adjacency matrix $A_G$ of an $d$-regular expander $G$. It always holds that the largest eigenvalue of $A_G$ in absolute value is $d$. Let $\lambda(G)$ denote the largest absolute value of any other eigenvalue. The following theorem is classical.
\begin{theorem}[\cite{FKS89}]\label{fact:graph_expander}
There exists absolute constants $c>1$ and $C>0$ such that for all sufficiently large $n$ and even $d$, there exists a $d$-regular $(n,n/2,c)$-expander $G$ such that $\lambda(G) \leq C\sqrt{d}$.
\end{theorem}

Next we present a result due to Upfal \cite{Upfal92}, implicitly used in the proof of Lemmas~1 and~2 therein. It states that there exists an expander graph of $n$ nodes and constant degree, such that after removing a constant fraction of nodes the remaining subgraph contains an expander of size $\Omega(n)$.
\begin{theorem}[\cite{Upfal92}]
\label{lem:graph_expander}
Let $G$ be an $(n,n/2,c)$-expander such that $G$ is $\delta$-regular and $\lambda(G) \leq C\sqrt{\delta}$, where $\delta$ is a (sufficiently large) constant and $c>1$, $C>0$ are absolute constants. There exist constants $\alpha,\zeta > 0$ and $\kappa > 1$, depending on $c$ and $C$, such that after removing an arbitrary set of at most $\zeta n$ nodes from $G$, the remaining graph contains a subgraph $G'$ such that $|V(G')| \geq \alpha n$ and $G'$ is a $(|V(G')|, n/2,\kappa)$-expander.
\end{theorem}

\subsection{Hashing schemes}

We employ two types of hashing schemes in our algorithm. To aid in the exposition of the analysis, it is useful to describe these in terms of their action on particular elements of a vector (i.e., they ``hash items into buckets''). The parameters $N, B_1, B_2, d_1, d_2$ of the hashing schemes are positive integers. We adopt the conventional notation that $[m] = \{1,2,\dots,m\}$.

\begin{definition}[one-layer hashing scheme]
The $(N,B,d)$ (one layer) hashing scheme is the uniform distribution on the set of all functions $f:[N]\to [B]^d$. We write $f(x) = (f_1(x),\dots,f_d(x))$, where $f_i$'s are independent $(N,B)$ hashing schemes.
\end{definition}
Each instance of such a hashing scheme induces a $d$-left-regular bipartite graph with $Bd$ right nodes. When $N$ is clear from the context, we simply write $(B,d)$ hashing scheme.
\begin{definition}[two-layer hashing scheme]
An $(N,B_1,d_1,B_2,d_2)$ (two-layer) hashing scheme is a distribution $\mu$ on the set of all functions $f:[N]\to [B_2]^{d_1d_2}$ defined as follows. Let $g$ be a random function subject to the $(N,B_1,d_1)$ hashing scheme and $\{h_{i,j}\}_{i\in[d_1],j\in[d_2]}$ be a family of independent functions subject to the $(B_1,B_2,d_2)$ hashing scheme which are also independent of $g$. Then  $\mu$ is defined to be the distribution induced by the mapping
\begin{multline*}
x\mapsto \left(h_{1,1}(g_1(x)),\dots,h_{1,d_2}(g_1(x)),h_{2,1}(g_2(x)),\dots,h_{2,d_2}(g_2(x)),\dots,\right.\\
\left.h_{d_1,1}(g_{d_1}(x)),\dots,h_{d_1,d_2}(g_{d_1}(x))\right).
\end{multline*}
\end{definition}
Each instance of such a hashing scheme gives a $d_1d_2$-left-regular bipartite graph of $B_2 d_1 d_2$ right nodes. When $N$ is clear from the context, we simply write $(B_1,d_1,B_2,d_2)$ hashing scheme. Conceptually we hash $N$ elements into $B_1$ buckets and repeat $d_1$ times, those buckets will be referred to as first-layer buckets; in each of the $d_1$ repetitions, we hash $B_1$ elements into $B_2$ buckets and repeat $d_2$ times, those buckets will be referred to as second-layer buckets.

We note that bipartite expander graphs can be used as hashing schemes because of their isolation property.

\begin{definition}[isolation property]
An $(n,m,d,\ell,\epsilon)$-bipartite expander $G$ is said to satisfy the $(\ell, \eta, \zeta)$-isolation property if for any set $S\subset L(G)$ with $|S|\leq \ell$, there exists $S'\subset S$ with $|S'|\geq (1-\eta)|S|$ such that for all $x\in S'$ it holds that $|\Gamma(\{x\})\setminus\Gamma(S\setminus\{x\})|\geq (1-\zeta) d$.
\end{definition}

\subsection{Bipartite expanders and hashing schemes}
\label{sec:hashing_and_expander}
All proofs use standard techniques and are postponed to the Appendix.

\subsubsection{One-layer Hashing}
\begin{lemma}[expanding property]\label{lem:one-layer}\label{LEM:ONE-LAYER}
For any $\epsilon \in (0, 1/4)$, $k\geq 1$, $\alpha\geq 1$ and $N = \Omega(\alpha k)$, a random one-layer $(B,d)$ hashing scheme gives an $(N,Bd,d,\alpha k,\epsilon)$-bipartite expander with probability $\geq 1-1/N^c$, where $B=\Omega(\frac{\alpha k}{\epsilon})$ and $d=\Omega(\frac{1}{\epsilon}\log\frac{N}{k})$.
\end{lemma}

\begin{lemma}[isolation property]\label{lem:one-layer-isolation}\label{LEM:ONE-LAYER-ISOLATION}
For any $\epsilon,\zeta \in (0, 1/4)$, $k\geq 1$, $\alpha\geq 1$ and $N = \Omega( k/\epsilon)$, a random one-layer $(B,d)$ hashing scheme gives a bipartite graph with $(L, \epsilon, \zeta)$-isolation property with probability $\geq 1-1/N^c$, 
where $B=\Omega(\frac{ k}{\zeta\epsilon})$, $d=\Omega(\frac{1}{\zeta\epsilon}\log\frac{N}{k})$, $L=O(k/\epsilon)$.
\end{lemma}

\subsubsection{Two-layer Hashing}
Now we show that a two-layer hashing scheme also gives a desirable bipartite expander. 

\begin{lemma}[expanding property]\label{lem:two-layer}\label{LEM:TWO-LAYER}
Let $\epsilon \in (0, 1/4)$, $k\geq 1$ and $N = \Omega(\max\{k/\epsilon^2, k^2\})$. A random two-layer $(B_1,d_1,B_2,d_2)$ hashing scheme gives an $(N,B_2d_1d_2,d_1d_2,4k,\epsilon)$-bipartite expander with probability $\geq 1-1/N^c$, where $B_1=\Omega(\frac{k}{\epsilon^2})$, $d_1=\Omega(\frac{1}{\epsilon}\frac{\log N}{\log(B_1/k)})$, $B_2 = \Omega(\frac{k}{\epsilon})$ and $d_2 = \Omega(\log\frac{B_1}{k})$ with appropriate choices of constants.
\end{lemma}
\begin{remark}\label{rem:constraint_k}
The constraint that $k = O(\sqrt{N})$ could be weakened to $k = O(N^{1-\xi})$ for any $\xi > 0$. The constants hidden in various $\Omega(\cdot)$ notations above will depend on $\xi$.
\end{remark}

We show that this two-layer hashing scheme also gives a good isolation property.
\begin{lemma}[isolation property]\label{lem:two-layer-isolation}\label{LEM:TWO-LAYER-ISOLATION}
Let $\epsilon > 0$, $\alpha>1$ be arbitrary constants and $(B_1,d_1,B_2,d_2)$ be a two-layer hashing scheme with $B_1=\Omega(\frac{k}{\zeta^\alpha\epsilon^{2\alpha}})$, $d_1=\Omega(\frac{\alpha}{\alpha-1}\cdot \frac{1}{\zeta\epsilon}\frac{\log N}{\log(B/k)})$, $B_2 = \Omega(\frac{k}{\zeta\epsilon})$ and $d_2 = \Omega(\frac{1}{\zeta}\log\frac{B_1}{k})$. Then with probability $\geq 1-1/N^c$, the two-layer hashing scheme with parameters prescribed above gives a bipartite graph with the $(L, \epsilon, \zeta)$-isolation property, where $L=O(k/\epsilon)$.
\end{lemma}


\section{Weak Recovery System}\label{sec:weak}
To simplify our analysis, we decompose a signal $\mb{x}$ into two parts of disjoint support, $\mb{x} = \mb{y} + \mb{z}$, where $\mb{y}$ has small support and $\mb{z}$ has small norm. By normalization, we may assume that $\|\mb{z}\|_1\leq 3/2$, where $3/2$ is chosen for the simplicity of constants in the proofs and can be replaced with an arbitrary positive number. We call $\mb{y}$ the \emph{head} and $\mb{z}$ the \emph{tail}. To simplify the language we may also use head to refer to $\supp{\mb{y}}$. We aim to recover the elements in $\mb{y}$. Introduced in \cite{PS12}, a \emph{weak system} takes an additional input, some set $I$ of indices (called the candidate set), and tries to estimate $\mb{x}_i$ for $i\in I$, hoping to recover some head items with estimate error dependent on $\|\mb{z}\|_1$. It is shown in \cite{PS12} that when $I$ contains the entire head, we can always recover a good fraction of the head. In this paper we make a slight modification on the definition of weak system as below. We only need $I$ to contain a good fraction of the head instead of the entire head.

\begin{definition}[Weak recovery system]
\label{def:weakI}
A \emph{Weak recovery system} consists of parameters $N,s,\eta,\zeta$, an $m$-by-$N$
{\em measurement matrix} $\mtx{\Phi}$, and a {\em decoding algorithm}
$\mathcal{D}$, that satisfy the following property:  

For any $\signal\in \mathbb{R}^N$ that can be written as
$\signal=\mathbf{y}+\mathbf{z}$, where $|\supp{\mathbf{y}}|\le s$ and
$\nerr{\mathbf{z}}_1\le 3/2$, given the
measurements $\mtx{\Phi}\signal$
and
a subset $I\subseteq[N]$ such that $|I\cap\supp{\mb{y}}|\geq (1-\zeta/2)|\supp{\mb{y}}|$, the
decoding algorithm $\mathcal{D}$ returns $\widehat\signal$, such that $\signal$ admits the following decomposition:
\[\signal=\widehat\signal+\widehat{\mathbf{y}}+\widehat{\mathbf{z}},\]
where

$|\supp{\widehat{\signal}}| = O(s)$,
$|\supp{\widehat{\mathbf{y}}}|\le \zeta s$, and 
$\nerr{\widehat{\mathbf{z}}}_1 \le \nerr{\mathbf{z}}_1+\eta$.
Intuitively, $\mb{\widehat y}$ and $\mb{\widehat z}$ will be the head and the tail of the residual $\signal-\widehat{\signal}$, respectively.
\end{definition}

\begin{algorithm}[bt]
\caption{Weak recovery system.}
\label{algo:weak}
\begin{algorithmic}
\Require{$N$, $s$, $\mtx{\Phi}$ (adjacency matrix of a $d$-left-regular expander $G$),
  $\mtx{\Phi}\signal$, and $I$}
\Ensure $\widehat\signal$
	\NFor{\textbf{each }$i\in I$}
    	\State $\signal_i' \gets \median_{u\in \Gamma(\{i\})} \sum_{(u,v)\in E} \mb{x}_u$
    	\Comment{each sum is an element of input $\mtx{\Phi}\signal$}
\State $\widehat\signal \gets $ top $O(s)$ elements of $\signal'$
\State \Return $\widehat\signal$
\end{algorithmic}
\end{algorithm}

\begin{theorem}[Weak Recovery]
\label{thm:weakI}
Suppose that $\Phi$ is the adjacency matrix of an $(N,Bd,d,4s,\eta)$-bipartite expander such that (a) $d=O(\frac{1}{\eta\zeta^2}\log\frac{N}{s})$ and $B = O(\frac{d}{\zeta\eta})$ and (b) it satisfies $(O(k/\eta),\eta,\zeta)$-isolation property. With appropriate instantiations of constants, Algorithm~\ref{algo:weak} yields a correct Weak recovery system that runs in time $O(|I|\eta^{-1}\zeta^{-2}\log(N/s))$.
\end{theorem}
The proof is essentially the same as \cite[Lemma 4]{PS12} but we hereby give a clearer abstraction by separating the deterministic argument from the randomized guarantees.

First, we need the following two lemmata.

\begin{lemma}[Noise]\label{lem:noise}
Let $\alpha>1$ and $t>\alpha s$. Let $\Phi$ be the adjacency matrix of an $(n,m,d,2\alpha s,\epsilon)$-bipartite expander with $\epsilon < 1/2$. Let $\mb{x}\in \mathbb{R}^n$ be such that $|\mb{x}_1|\geq |\mb{x}_2|\geq \cdots \geq |\mb{x}_n|$.  Let $I = \{1,\dots,\alpha s\}$, then 
\[
\|(\Phi(\mb{x}-\mb{x}_{[t]}))_{\Gamma(I)}\|_1 \leq 4\epsilon d(\|\mb{x}-\mb{x}_{[t]}\|_1 + \alpha s |\mb{x}_{t+1}|).
\]
\end{lemma}
\begin{proof}
Partition $\{1,\dots,N\}$ into blocks $I\cup H_1\cup B_1\cup B_2\cup \dots$, where $H_1=\{\alpha s+1,\dots,t\}$ and $B_i=\{t+(i-1)\alpha s+1,\dots,t+i\alpha s\}$ for $i\geq 1$. Consider $\mb{x}$ restricted to a block $B_i$. We consider the following two cases.

\textbf{Case 1}. $\mb{x}_{B_i}$ is quasi-flat, i.e., $|\mb{x}_{t+i\alpha s}|\geq |\mb{x}_{t+(i-1)\alpha s+1}|/2$. (It is called quasi-flat because all entries are within a factor of $2$ of each other.) Consider all $d|B_i|$ edges in the expander emanating from $B_i$. Suppose that $Z$ edges of them are incident to $\Gamma(I)$, then
\[
|\Gamma(I)\cup \Gamma(B_i)|\leq d(|I|+|B_i|) - Z.
\]
On the other hand, by the expansion property,
\[
|\Gamma(I)\cup \Gamma(B_i)|\geq (1-\epsilon) d(|I|+|B_i|),
\]
which implies that
\[
Z\leq \epsilon d(|I|+|B_i|)\leq 2\epsilon\alpha kd.
\]
It follows that
\[
\|(\Phi \mb{x}_{B_i})_{\Gamma(I)}\|_1\leq Z\cdot \max_{i\in B_i}|\mb{x}_i|\leq 2\epsilon\alpha kd\cdot |\mb{x}_{t+(i-1)\alpha s+1}|\leq 4\epsilon d\|\mb{x}_{B_i}\|_1,
\]
where the last inequality follows from the fact that $\mb{x}_{B_i}$ is quasi-flat so that $\alpha s|\mb{x}_{t+(i-1)\alpha s+1}|\leq 2\|\mb{x}_{B_i}\|_1$.

\textbf{Case 2}. $\mb{x}_{B_i}$ is not quasi-flat, then $|\mb{x}_{t+i\alpha s}| < |\mb{x}_{t+(i-1)\alpha s+1}|/2$. Let 
\[
J = \{i\in B_i: |\mb{x}_i| < |\mb{x}_{t+(i-1)\alpha s+1}|/2\}.
\]
Increase $|\mb{x}_i|$ for all $i\in J$ so that $|\mb{x}_i| = |\mb{x}_{t+(i-1)\alpha s+1}|/2$ and $\mb{x}_{B_i}$ becomes flat, and this increases $\|\mb{x}_{B_i}\|_1$ by at most $\alpha s |\mb{x}_{t+(i-1)\alpha s+1}|/2$. Invoking Case 1, we obtain that
\[
\|(\Phi \mb{x}_{B_i})_{\Gamma(I)}\|_1 \leq 4\epsilon d\left(\|\mb{x}_{B_i}\|_1 + \frac{\alpha s \left|\mb{x}_{t+(i-1)\alpha s+1}\right| }{2}\right).
\]
Now we go back to the entire $\mb{x}$. Suppose that $\mb{x}_{B_{i_1}},\dots,\mb{x}_{B_{i_q}}$ are not quasi-flat, then by triangle inequality we shall have
\[
\|(\Phi(\mb{x}-\mb{x}_t))_{\Gamma(I)}\|_1 \leq 4\epsilon d\|\mb{x}-\mb{x}_t\|_1 + 4\epsilon d\cdot \frac{\alpha s}{2}\sum_{p=1}^q \left|\mb{x}_{t+(i_p-1)\alpha s+1}\right|.
\]
Observe that for $p\geq 2$,
\[
|\mb{x}_{t+(i_p-1)\alpha s+1}|\leq |\mb{x}_{t+i_{p-1}\alpha s+1}| \leq \frac{|\mb{x}_{t+(i_{p-1}-1)\alpha s+1}|}{2},
\]
where the last inequality follows from the quasi-flatness of $\mb{x}_{B_{i_{p-1}}}$. Hence,
\[
|\mb{x}_{t+(i_p-1)\alpha s+1}|\leq \frac{|\mb{x}_{t+1}|}{2^{p-1}},\quad p\geq 1,
\]
whence it follows that
\[
\|(\Phi(\mb{x}-\mb{x}_{[t]}))_{\Gamma(I)}\|_1 \leq 4\epsilon d(\|\mb{x}-\mb{x}_{[t]}\|_1 + \alpha s |\mb{x}_{t+1}|).
\]
\end{proof}
In the usual decomposition, the head contains the entries with large coordinate values, which will be referred to as \emph{heavy hitters}. If a heavy hitter fails to be recovered, it must have been displaced by another entry, called a decoy, in the recovered signal. The next lemma bounds the number of decoys.
\begin{lemma}[Decoys]\label{lem:decoy}
Let $\theta, \epsilon\in (0,1)$ and $\beta,\zeta > 0$ such that $0 < \zeta < \frac12 - \frac{80\beta}{\theta}$.
Suppose that $G$ is a $(4s,d,\beta \epsilon)$-bipartite expander which satisfies the $(\frac{9s}{\epsilon},\frac{\epsilon\theta}{18},\zeta)$-isolation property. Let $\mb{x}\in \mathbb{R}^n$ be a signal satisfying the assumption in the Weak system, and let $\mb{x}'\in \mathbb{R}^n$ be the estimates defined as
\[
\mb{x}'_{i} = \median_{u\in \Gamma(\{i\})} \sum_{(u,v)\in E} \mb{x}_u,\quad i\in [N].
\]
Define
\[
D = \{i\in \supp{\mb{y}}: |\mb{x}_i - \mb{x}'_i| \geq \epsilon/(4s)
\},
\]
then $|D| < \theta s$.
\end{lemma}
\begin{proof}
Suppose that $|D|\geq \theta s$. By definition it holds that $|D|\leq s$. Also assume that $|\mb{x}_1|\geq |\mb{x}_2|\geq \cdots \geq |\mb{x}_n|$. Suppose that $|\mb{x}_i|\geq \epsilon/(2s)$ for all $i\in H:=\supp{\mb{y}}$, otherwise we can place the violated $i$'s into $\mb{z}$, causing $\|\mb{z}\|_1$ to increase by at most $s\cdot \epsilon/(2s) = \epsilon/2$, so we would have $\|\mb{z}\|_1\leq 2$. Let $T = H\cup D\cup \{i: |\mb{x}_i|\geq \epsilon/(4s)\}$, then $t := |T|\leq \|\mb{z}\|_1/(\epsilon/(4s)) + |D| + |H|\leq 9s/\epsilon$. 

Note that $|\mb{x}_{t+1}| \leq \epsilon/(4s)$. Taking $\alpha = 2$ in Lemma~\ref{lem:noise}, we know that
\[
\|(\Phi(\mb{x}-\mb{x}_{[t]}))_{\Gamma(H\cup D)}\|_1 \leq 4\cdot\beta\epsilon d\left(\frac{3}{2}+ \frac{\epsilon}{2} + 2s\cdot\frac{\epsilon}{4s}\right) \leq 10\beta\epsilon d.
\]
By the isolation property, there are at most $\frac{9s}{\epsilon}\cdot \frac{\epsilon\theta}{18} = \frac{\theta s}{2}$ elements in $T$ which are not isolated in at least $(1-\zeta) d$ nodes from other elements in $T$. This implies that at least $\theta s/2$ elements in $D$ are isolated in at least $(1-\zeta) d$ nodes from other elements in $T$.

A decoy at position $i$ receives at least $\epsilon/(4s)$ noise in at least $(1/2-\zeta) d$ isolated nodes of $\Gamma(\{i\})$, hence in total, a decoy element receives at least $\epsilon(1/2-\zeta)d/(4s)$ noise. Therefore the $\theta s/2$ decoys overall should receive noise at least
\[
\frac{\epsilon(\frac12-\zeta)d}{4s}\cdot \frac{\theta s}{2} > 10\beta\epsilon d\geq \|(\Phi(x-x_t))_{\Gamma(H\cup D)}\|_1,
\]
which is a contradiction. Therefore $|D| < \theta s$.
\end{proof}
Now we are ready to show Theorem~\ref{thm:weakI}.
\begin{proof}[Proof of Theorem~\ref{thm:weakI}]
The proof is essentially the same as \cite[Lemma 4]{PS12}. It follows from Lemma~\ref{lem:decoy} that with appropriate choices of constants, that there are at most $\zeta s/4$ decoys and at least $(1-\zeta/4)s$ elements $i$ in $\supp{\mb{y}}$ satisfy $|\mb{x}_i - \mb{x}_i'|\leq \eta/(4s)$. Let $I' = I\cap \supp{\mb{y}}$. We describe below the construction of $\widehat{\mb{x}}$, $\widehat{\mb{y}}$ and $\widehat{\mb{z}}$.
\begin{itemize}
\item Elements $i\in\supp{\widehat\signal}$ with a good estimate (to
  within $\pm \eta/(4s)$ contribute $\signal_i-\widehat\signal_i$ to
  $\widehat{\mathbf{z}}$.  There are at most $s$ of these, each
  contributing $\eta/(4s)$, for total contribution $\eta/4$ to
  $\widehat{\mathbf{z}}$.
\item Elements $i\in\supp{\widehat\signal}$ with a bad estimate (not
  to within $\pm \eta/(4s)$) contribute $\signal_i-\widehat\signal_i$ to
  $\widehat{\mathbf{y}}$.  There are at most $\zeta s/4$ of these.
\item Elements $i\in\supp{\mathbf{z}}\setminus\supp{\widehat\signal}$
  contribute $\signal_i$ to $\widehat{\mathbf{z}}$.  The $\ell_1$ norm of these is  at most $\|\mathbf{z}\|_1$.
\item Elements $i\in I'\setminus\supp{\widehat\signal}$
  with a good estimate that are nevertheless displaced by another
  element $i'\in\supp{\widehat\signal}\setminus\supp{\mathbf{y}}$ with
  a good estimate contribute to $\widehat{\mathbf{z}}$.
  There are at most $s$ of these.  While the value $\signal_i$ may be
  large and make a large contribution to $\widehat{\mathbf{z}}$, this
  is offset by $\signal_{i'}$ satisfying
$|\signal_{i'}|
\ge |\widehat{\signal}_{i'}|-\eta/(4s)
\ge |\widehat{\signal}_{i}|-\eta/(4s)
\ge |\signal_{i}|-\eta/(2s)$, which 
  contributes to $\mathbf{z}$ but not to
  $\widehat{\mathbf{z}}$.  Thus the net contribution to
  $\widehat{\mathbf{z}}$ is at most $\eta/(2s)$ for each of the $s$
  of these $i$, for
  a total $\eta/2$ contribution to $\widehat{\mathbf{z}}$.

\item Elements $i\in I'\setminus\supp{\widehat\signal}$
  that themselves have bad estimates or are displaced by elements with
  bad estimates contribute $\signal_i$ to $\widehat{\mathbf{y}}$.  There are at
  most $\zeta s/4$ bad estimates overall, so there are at most $\zeta s/4$
  of these.
\item Elements $i\in I\setminus I'$ contribute to $\widehat{\mb{y}}$. There are at most $\zeta s/2$ of these.
\end{itemize}
It is clear that $|\supp{\widehat{\mb{y}}}|\leq \zeta s$ and $\|\widehat{\mb{z}}\|_1\leq \|\mb{z}\|_1 + \eta$, as desired. The runtime is easy to verify.
\end{proof}
To complete the construction of a weak recovery system, we refer the reader to Section~\ref{sec:hashing_and_expander} to show that a bipartite expander as required by Theorem~\ref{thm:weakI} exists. We show, by probabilistic methods, that it can be attained by both one-layer and two-layer hashing schemes, with appropriate parameters. For example, if we combine Lemma~\ref{lem:one-layer}, Lemma~\ref{lem:one-layer-isolation}, and Theorem~\ref{thm:weakI}, we have a clean formulation, in the language of expanders, of the result on weak system in \cite{PS12}.

\section{Identification of Heavy Hitters}\label{sec:backpointers}
In the previous section, we showed how to estimate all candidates in a candidate set $I$ quickly. The main bottleneck in a highly efficient algorithm is finding a non-trivial set $I \subset [N]$ of candidates which we address in this section.

The overall strategy is as follows. Using the two-layer hashing scheme $(B_1,d_1,B_2,d_2)$, we expect that a heavy hitter dominates the first-layer buckets where it lands in $\Omega(d_1)$ repetitions. In each of these repetitions, it is a heavy hitter in a signal of length $B_1$, and we expect to recover it using the Weak algorithm applied to the signal of length $B_1$ with $I = [B_1]$. After finding the heavy buckets in each repetition, the remaining problem is to extract the position of a heavy hitter $i$ from the $\Omega(d_1)$ repetitions that contain $i$. 
Recall, as previewed in the introduction, that we shall assign to each index $i\in [N]$ a message $\mb{m}_i$, which uniquely identifies the index $i$. The message will be encoded in the measurement matrix $\bm{\Phi}$ and we expect to recover the message $\mb{m}_i$ for heavy hitters $i$ from the measurements $\bm{\Phi}\mb{x}$ and thus the index $i$. The recovery of the message is to be done block by block, which motivates the following definition of {\em Weak List Recovery Criterion}.
\begin{definition}[Weak List Recovery Criterion]
\label{def:criterionI}
Fix $N,s$. Suppose that $\signal\in \mathbb{R}^N$ can be written as
$\signal=\mathbf{y}+\mathbf{z}$, where $|\supp{\mathbf{y}}|\le s$ and
$\nerr{\mathbf{z}}_1\le 3/2$. 
Let $\mb{m} = (\mb{m}_1,\dots,\mb{m}_N)$, where each $\mb{m}_i$ is a binary string (also called a \emph{message}) of length $\beta$. Suppose $\widehat{\mb{m}}$ is a
list of possible index-message pairs, that is, $\widehat{\mb{m}}\subseteq [N]\times \{0,1\}^\beta$. We say that
$\widehat{\mb{m}}$ is {\em correct in the weak list recovery sense} if $(i,\mb{m}_i)\in\widehat{\mb{m}}$ for at least $|\supp{\mb{y}}|-s/8$ indices $i$ in $\supp{\mb{y}}$.
\end{definition}
The encoding/decoding scheme is given in Algorithm~\ref{algo:encoding_paradigm}. We break each message $\mb{m}_i$ associated with position $i$ into $d_1$ blocks, $\mb{m}_{i,1},\dots,\mb{m}_{i,d_1}$. Note that $\mb{m}_i$ could be much longer than $\log N$ bits in order to guarantee a successful list recovery. Now in the $j$-th repetition of the $d_1$ repetitions, we obtain a signal $\widetilde{\mb{x}}$ of length $B$. Each $\widetilde{\mb{x}}_\ell$ is associated with a message that can be viewed as a weighted sum of $\mb{m}_{i,j}$ for positions $i$ hashed into bucket $\ell$. If a heavy hitter $i$ is isolated in bucket $\ell$ and the noise is mild in this bucket, this weighted sum would be approximately $\mb{m}_{i,j}$, and we expect to recover $\mb{m}_{i,j}$ from the second-layer hashing, with inner encoding and decoding. Now we assume that we have recovered $\mb{m}_{i,j}$ for heavy hitter $i$ in sufficiently many repetitions $j$. The central difficulty is to match $\mb{m}_{i,j}$ with $\mb{m}_{i,j'}$ with $j\neq j'$ in order to find enough fraction of $\mb{m}_i$ in the end. In order to solve this we shall encode some linking information in the node that will enable us to match $\mb{m}_{i,j}$ with $\mb{m}_{i,j'}$. This will be the topic of the next subsection, in which we shall use the Parvaresh-Vardy code to overcome this difficulty.
\begin{algorithm}[bt]
{\small
\caption{Encding/Decoding paradigm.}
\label{algo:encoding_paradigm}
\begin{algorithmic}
\State{// Encoding with $(B_1,d_1,B_2,d_2)$ hashing scheme}
\For{ $i=1$ to $N$}
  \State{{\bf Break}: Break the information of $i$ into
    $d_1$ blocks}
  \State{{\bf Outer encoding}: Encode the blocks with cluster info (from a regular expander graph) and against errors, getting $\{\mb{m}_{i,j}\}_{j=1}^{d_1}$}
\EndFor
\For{ $j=1$ to $d_1$}
  \State{{\bf Inner encoding}: Encode $\mb{m}_{i,j}$, for $i\in[N]$}
\EndFor
\State{// Decoding with $(B_1,d_1,B_2,d_2)$ hashing scheme}  
\For{ $j=1$ to $d_1$}
  \State{{\bf Inner decoding}: Recover $\widehat{\mb{m}}_j$ in the Weak List sense}
  \State{{\bf Record Side Info}: Tag each element of $\widehat{\mb{m}}_j$ with $j$}
\EndFor
\State{{\bf Outer decoding}: From $\widehat{\mb{m}}=\bigcup_j
  \widehat{\mb{m}}_j$'s, find
  block clusters and correct errors; produce $I$}
\end{algorithmic}}
\end{algorithm}

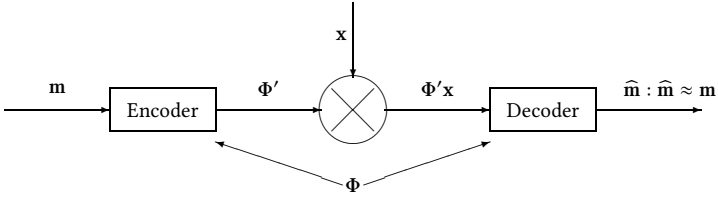
\begin{figure}
\begin{center}
\scalebox{.8}{
\begin{picture}(330,85)(0,20)

\put( 50, 40){\framebox(50,20)[c]{Encoder}}

\put(165,15){\makebox(0,0)[c]{${\mtx\Phi}$}}
\put(160,15){\vector(-3,1){60}}

\put(170,15){\vector( 3,1){60}}

\put(230, 40){\framebox(50,20)[c]{Decoder}}

\put(  0, 50){\vector(1,0){50}}
\put( 25, 60){\makebox(0,0)[c]{$\mb{m}$}}

\put(100, 50){\vector(1,0){50}}
\put(125, 60){\makebox(0,0)[c]{$\mtx\Phi'$}}

\put(165, 50){\circle{30}}

\put(165, 50){\line(+1,+1){10}}
\put(165, 50){\line(+1,-1){10}}
\put(165, 50){\line(-1,+1){10}}
\put(165, 50){\line(-1,-1){10}}

\put(165,100){\vector(0,-1){35}}

\put(162,85){\makebox(0,0)[r]{$\mb{x}$}}

\put(180, 50){\vector(1,0){50}}
\put(205, 60){\makebox(0,0)[c]{$\mtx\Phi' \mb{x}$}}

\put(280, 50){\vector(1,0){50}}
\put(315, 60){\makebox(0,0)[c]{$\widehat{\mb{m}}:\widehat{\mb{m}}\approx \mb{m}$}}
\end{picture}
}
\end{center}
\caption{The encoder and decoder agree on some matrix $\mtx\Phi$. The encoder takes messages $\mb{m}$ and produces a measurement matrix $\mtx\Phi'$ based on $\mb{m}$ and $\mtx\Phi$. The system takes input $\mb{x}$ and produces measurements $\mtx\Phi'\mb{x}$, from which the decoder tries to recover $\widehat{\mb{m}}$ in the sense of weak list recovery.}
\label{fig:channel}
\end{figure}

Next we illustrate our idea of encoding with a simple case of the sparse recovery problem, where we wish to find $k$ heavy hitters among $B$ positions. We shall encode messages with length $\beta = \log(B/k)$ using $k \log (B/k)$ measurements and recover the messages associated with $\Omega(k)$ heavy hitters in time approximately $B$. To better illustrate the idea, we refer the reader to Figure~\ref{fig:channel}.
\begin{lemma}
\label{lem:R-S_coding}
Fix $k$, $B$, $\epsilon$, where $B=\Omega(k/\epsilon)$ and consider the sparse recovery problem of finding $k$ heavy hitters among $B$ positions. Let $\beta = O(\log(B/k))$. There is a coding scheme to encode messages of length $\beta$ using $m=O((k/\epsilon)\log(B/k))$ measurements and recover the messages in the weak list recovery sense with decoding running in
time $O(B\log^3(B/k))$. This scheme also uses a look up table of size $\beta$.
\end{lemma}
\begin{proof}
As an outer code, use Reed-Solomon over an alphabet of size
$\beta/\log \beta$.  This is concatenated with a random code of length
$\log\beta$ as an inner code.  The inner code can be decoded in
constant time from a lookup table of size $\beta$ and the
outer code can be decoded by solving a linear system of size
approximately $\beta$ in time $O(\beta^2)$. 
Hence for each index $i\in [B]$, the message $\mb{m}_i$ of length $\beta$ is encoded into a longer message $\mb{m}_i'$ of length $\beta'$, where $\beta' = \Theta(\beta)$. It suffices to demonstrate how to encode and decode the longer messages $\mb{m}_i'$.

We use a Weak system (Theorem~\ref{thm:weakI}) with a $(\Theta(k),d,\epsilon)$-bipartite expander that exhibits a $(\Theta(k),d)$ hashing scheme and satisfies $(O(k/\epsilon), \epsilon, O(1))$-isolation property, where $d=\Theta(\log(B/k))\geq \beta'$. Let $\mtx{\Phi}$ be the adjacency matrix of the bipartite expander. Without loss of generality, we may assume that $\beta' = d$.

Now we describe the construction of the new measurement matrix $\mtx{\Phi}'$, which has twice as many rows as $\mtx{\Phi}$. Viewing $\mtx{\Phi}$ as a hashing matrix with $\beta'$ repetitions. For each $i\in [B]$ and $j\in [\beta']$, we need to encode the $j$-th bit of the messages $\mb{m}_i'$ in the $j$-th repetition. As there are $\beta'$ repetitions, a total of $\beta'$ bits will be encoded for each index $i\in [B]$, as desired. For each row $\rho$ of $\mtx{\Phi}$ in the $j$-th repetition of hashing, we construct a $2\times N$ submatrix $\rho'$ as follows. For each $i\in [B]$, the $i$-th column of $\rho'$ is 
$\bigl(\begin{smallmatrix}\rho_i\\0\end{smallmatrix}\bigr)$ when $\mb{m}_{i,j}' = 1$ and
$\bigl(\begin{smallmatrix}0\\\rho_i\end{smallmatrix}\bigr)$ when $\mb{m}_{i,j}' = 0$. Note that either is a column of two zeroes when $\rho_i = 0$. In this way each row of $\mtx{\Phi}$ induces two rows of $\mtx{\Phi}'$.

Finally we show how to recover the messages. To decode one bit,
consider any
$\bigl(\begin{smallmatrix}a\\b\end{smallmatrix}\bigr)$
to be a {\em relaxed} encoding equivalent to
$\bigl(\begin{smallmatrix}\rho_i\\0\end{smallmatrix}\bigr)$
if $|a|>|b|$ and 
$\bigl(\begin{smallmatrix}0\\\rho_i\end{smallmatrix}\bigr)$ otherwise, where $\rho$ is a row of $\mtx{\Phi}$. We know that there exist $\Omega(k)$ heavy hitters, each dominates the buckets where it lands in $\Omega(d)$ repetitions. In each such repetition, our bit encoding scheme ensures that the associated bit can be recovered successfully, hence for each of such heavy hitter, we shall collect $\Omega(d)$ bits, enough to recover the encoded message of $\beta'$ bits and thus the original message of $\beta$ bits (using Theorem~\ref{thm:weakI} for the weak system with $I = [B]$).

The runtime is $O(B\beta^2\log(B/k))$ for exhaustive recovery in the Weak system.
\end{proof}

\begin{remark} In the proof above, the matrix $\mtx\Phi$ is not necessarily the matrix of a one-layer hashing scheme. If $\Phi$ is a `layer-structured' matrix, the same row-doubling construction $\mtx\Phi'$ can be employed. For the one-layer $(B,d)$ hashing scheme, the matrix $\mtx\Phi$ can be viewed as having $B$ layers, each corresponds to a repetition. For the two-layer $(B_1,d_1,B_2,d_2)$-hashing scheme, the matrix $\mtx\Phi$ can be viewed as having $B_1 B_2$ layers (each layer is a repetition in the second-layer hashing). This observation will be used in our main construction (Lemma~\ref{lem:expander_recovery}).
\end{remark}

\begin{remark} The Reed-Solomon code is used in the proof. In general, any code that has a constant rate and constant error radius and can be decoded in linear time (up to polylogarithmic factors) will work. The decoding runtime in the lemma statement will be adjusted accordingly.
\end{remark}

\begin{figure}
      \begin{minipage}[b]{0.45\linewidth}
\begin{tikzpicture}[scale=0.8,every node/.style={transform shape}]
\foreach \y/\j in {0/1,1.2/2,2.4/3,4.5/4}{
\foreach \i in {0,...,4}{
\node[gray,fill,circle,inner sep=2pt,minimum size=5pt,outer sep=2pt] (a\j\i) at (\y, 4+\i) {};
};
\draw(\y,3.3) node {$\vdots$};
\node[gray,fill,circle,inner sep=2pt,minimum size=5pt,outer sep=2pt] (b\j)at (\y,2.5) {};
\draw[arrows={-latex},thick,bend right=50] (\y-0.1,2.5+0.9) to (b\j);

\draw[arrows={-latex},thick] (a\j4) to (a\j3);
\draw[arrows={-latex},bend left,thick] (a\j4) to (a\j2);
\draw[arrows={-latex},thick] (a\j3) to (a\j2);
\draw[arrows={-latex},bend left,thick] (a\j2) to (a\j0);
\draw[arrows={-latex},bend right=25,thick] (a\j4) to (a\j1);

};
\draw(3.5,5) node {$\cdots$};
\node[anchor=base] (text1) at (2.4,8.5) {$N$ columns};
\node[anchor=base] (text2) at (-1.5,5.5) {$d_1$ rows};
\draw[arrows={-latex},thick] (text1) to (0,8.6);
\draw[arrows={-latex},thick] (text1) to (4.5,8.6);
\draw[arrows={-latex},thick] (text2) to (-1.5,8);
\draw[arrows={-latex},thick] (text2) to (-1.5,2.5);
\draw(0,2) node {$\mb{x}_1$};
\draw(1.2,2) node {$\mb{x}_2$};
\draw(2.4,2) node {$\mb{x}_3$};
\draw(4.5,2) node {$\mb{x}_N$};
\end{tikzpicture}
              \caption{Underlying graph $G_N$. Suppose that $\mb{x}_1$ is in the tail and that $\mb{x}_2$, $\mb{x}_3$ and $\mb{x}_N$ are heavy hitters.}
				\label{fig:underlying_graph}
      \end{minipage}
      \hspace{0.05\linewidth}
      \begin{minipage}[b]{0.45\linewidth}
      \centering

\begin{tikzpicture}[scale=0.8,every node/.style={transform shape}]
\foreach \y/\j in {0/1,1.2/2,2.4/3,4.5/4}{
\foreach \i in {0,...,4}{
\node[gray,fill,circle,inner sep=2pt,minimum size=5pt,outer sep=2pt] (a\j\i) at (\y, 4+\i) {};
};
\draw(\y,3.3) node {$\vdots$};
\node[gray,fill,circle,inner sep=2pt,minimum size=5pt,outer sep=2pt] (b\j)at (\y,2.5) {};
\node[circle] (hidden\j) at (\y-0.1,2.5+0.9) {};
};
\foreach \j in {2,3,4}{
\draw[arrows={-latex},thick,bend right=50] (hidden\j) to (b\j);

\draw[arrows={-latex},thick] (a\j4) to (a\j3);
\draw[arrows={-latex},bend left,thick] (a\j4) to (a\j2);
\draw[arrows={-latex},thick] (a\j3) to (a\j2);
\draw[arrows={-latex},bend left,thick] (a\j2) to (a\j0);
\draw[arrows={-latex},bend right=25,thick] (a\j4) to (a\j1);
};
\draw(3.5,5) node {$\cdots$};
\draw[arrows={-latex},thick,bend right=10](a13) to (a21);
\draw[arrows={-latex},thick,bend right] (3.3,5.5) to (a33);
\draw[arrows={-latex},thick,bend left] (3.7,6) to (a42);
\draw[arrows={-latex},thick,out=180,in=-75] (3.4,3) to (a20);
\end{tikzpicture}
              \caption{Recovered graph $\tilde G$ in ideal situation, with expander copies clairvoyantly aligned in a column. Since the first column corresponds to a tail item and is thus not expected to be recovered, it is almost absent in the recovered graph. For each heavy-hitter column, the whole copy of the expander graph is expected to be recovered. There may exist some arcs from a non-heavy-hitter column to a heavy-hitter column, but not vice versa.}
				\label{fig:ideal_recovery}
      \end{minipage}
\end{figure}

\begin{figure}
\centering
\begin{tikzpicture}[scale=0.8,every node/.style={transform shape}]
\foreach \y/\j in {0/1,1.2/2,2.4/3,4.5/4}{
\foreach \i in {0,...,4}{
\node[gray,fill,circle,inner sep=2pt,minimum size=5pt,outer sep=2pt] (a\j\i) at (\y, 4+\i) {};
};
\draw(\y,3.3) node {$\vdots$};
\node[gray,fill,circle,inner sep=2pt,minimum size=5pt,outer sep=2pt] (b\j)at (\y,2.5) {};
\node[circle] (hidden\j) at (\y-0.1,2.5+0.9) {};
};
\foreach \j in {2,3,4}{
\draw[arrows={-latex},thick,bend right=50] (hidden\j) to (b\j);
};
\draw (a24) circle(0.15);
\draw (a34) circle(0.15);
\draw (a44) circle(0.15);

\draw(3.5,5) node {$\cdots$};
\draw(0,2) node {$\mb{x}_1$};
\draw(1.2,2) node {$\mb{x}_2$};
\draw(2.4,2) node {$\mb{x}_3$};
\draw(4.5,2) node {$\mb{x}_N$};

\draw[arrows={-latex},thick,bend right=10](a13) to (a21);
\draw[arrows={-latex},thick,in=45,out=180](a24) to (a11);
\draw[arrows={-latex},thick,in=135,out=-30](a24) to (a43);
\draw[arrows={-latex},thick,bend left](a24) to (a22);
\draw[arrows={-latex},thick](a23) to (a22);
\draw[arrows={-latex},thick](a33) to (a22);
\draw[arrows={-latex},thick,bend left](a22) to (a20);
\draw[arrows={-latex},thick,bend left](a34) to (a32);
\draw[arrows={-latex},thick,bend left](a32) to (a30);
\draw[arrows={-latex},thick,bend right](a34) to (a31);
\draw[arrows={-latex},thick](a34) to (a33);
\draw[arrows={-latex},thick](a43) to (a42);
\draw[arrows={-latex},thick,bend left](a42) to (a40);

\draw[arrows={-latex},thick,bend right] (3.3,5.5) to (a33);
\draw[arrows={-latex},thick,bend left] (3.7,6) to (a42);
\draw[arrows={-latex},thick,out=180,in=-75] (3.4,3) to (a20);
\end{tikzpicture}
\caption{Recovered graph $\tilde G$, with `supposed' expander copies clairvoyantly aligned in columns. The first column corresponds to a tail item so it is almost absent. The top node in the second column is corrupted so it points to wrong columns but nevertheless the correct rows because the row information is hard-wired. The top node in the third column is correctly recovered but the second node in the column is corrupted. The top node in the last column has a small bucket value in the first repetition so it is absent $\tilde G$. If we perform BFS at the top node in the third column, we may include a lot of nodes in the second column.}
\label{fig:real_recovery}
\end{figure}
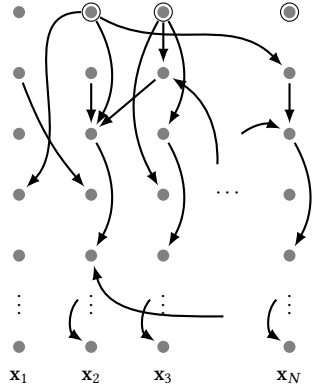

\subsection{Expander Encoding}
\label{sec:expander}

\label{sec:expander_description}

\noindent\textbf{Parameters} We assume that the constants $\beta > 0$ and $\gamma \in (0,1)$ are fixed; the parameters $B_1$, $d_1$, $B_2$, $d_2$ are as in Lemma~\ref{lem:two-layer-isolation} such that 
$B_1 = \Theta\bigl((\frac{k}{\epsilon^2})^{1+\beta}\log\frac{N}{k}\bigr)$ 
and $d_1 = \big(\frac{1}{\epsilon} + \big(\frac{\log N}{\log k}\big)^\gamma\big)\frac{\log N}{\log(B_1/k)} $; 
$c \leq m$ are constant integers; $h$ is an integer; and $\epsilon = O\big( \big(\frac{\alpha}{m}\big)^\frac{m}{m-c}\big(\frac{\log(B_1/k)}{\log N}\big)^{\gamma}\big)$. 

Let $G$ be a graph of $d_1$ nodes with constant degree $\delta$ that satisfies Theorem~\ref{fact:graph_expander}, and $\alpha,\zeta,\kappa$ be constants provided by Theorem~\ref{lem:graph_expander} when applied to $G$. Without loss generality we can assume that $\alpha\leq 1/2$. Adjust the hidden constants together with $c$, $m$ and $h$ appropriately (depending on $\beta$ and $\gamma$) such that
\begin{enumerate}
\renewcommand{\labelenumi}{(\alph{enumi}) }
\setlength{\itemsep}{0pt}
\setlength{\parskip}{0pt}
\setlength{\parsep}{0pt}  
    
	\item $B_1 > d_1$;
	\item $(h-1) m\log_{B_1} N  < \alpha d_1$;
	\item $(\alpha d_1 - (h-1)m\log_{B_1} N)\cdot h^m > d_1^c$;
	\item $c \geq \log\delta/\log\kappa$.
\end{enumerate} 
We note that an instance of $m,h$ is to choose $m\geq c(1+1/\gamma)$ and $h = \Theta(d_1^{c/m})$.

\paragraph{Encoding} We shall use Reed-Solomon for inner encoding.  Next, we define our outer coding, which uses the Parvaresh-Vardy code \cite{PV05}. Take $N$ disconnected copies of $G$ and call the union $G_N$, where each node is indexed by a pair $(i,r)\in [N]\times [d_1]$. See Figure~\ref{fig:underlying_graph}. Also, let $\F$ be a field such that $|\F| = \Theta(B_1)$ is a power of $2$ and $E(x)$ be an irreducible monic polynomial over $\F$ such that $\deg E(x) = \log_{B_1}N$. View each $i\in [N]$ as a polynomial $f$ over $\F$ with degree $\log_{B_1} N - 1$. For each $(i,r)\in G_N$, associate with it an element $p(i,r) \in \F^{m+1}$ as
\[
p(i,r) = (x_{i,r}, f(x_{i,r}), (f^h \bmod{E})(x_{i,r}),\dots, (f^{h^{m-1}}\bmod{E})(x_{i,r})),
\]
where $f$ is a polynomial associated with $i\in [N]$ and $x_{i,r}\in \F$ so that $x_{i,r}$ are distinct for different $r$. This is possible because of Property (a).

Attach to a node $(i,r)$ a message $\mb{m}_{i,r}$ containing the information of $p(i,r)$ as well as $H(i, v_1(r))$,$\dots$, $H(i,v_\delta(r))$, where $v_1(r),\dots,v_\delta(r)$ are the neighbours of $r$ in $G$ and $H(i,j)\in [B_1]$ gives the bucket index where $i$ lands in the $j$-th outer hashing repetition.  It is clear that $\mb{m}_{i,r}$ has $\Theta(\log B_1) = O(d_2)$ bits and therefore we can encode it in $d_2$ hash repetitions, see Lemma~\ref{lem:R-S_coding}.

\paragraph{Decoding} In each of the $d_1$ repetitions, we shall recover $O(k/\epsilon)$ heavy buckets and thus obtain $O(k/\epsilon)$ nodes with their messages. Even when the messages are recovered correctly, we only know that a message corresponds to $\mb{m}_{i,r}$ for some $i\in [N]$ and we do not know which $i$ it is. However, if we can determine that enough messages are associated with the same $i$, we would have obtained enough $p(i,r)$ for different values of $r$ then we should be able to find $f$ and thus recover the position $i$.

To determine enough $p(i,r)$ for the same $i$, we do clustering as follows. Suppose that there are $k$ heavy hitters at position $i_1,\dots,i_k$. Let $\widetilde G$ be a graph of $d_1\times O(k/\epsilon)$ nodes, arranged in a $d_1\times O(k/\epsilon)$ grid. For now we assume that the messages are recovered correctly for each heavy hitter $i$ in all $d_1$ repetitions. (This means that there are no collisions and the noise in the buckets are all small.)
Each message has the form $p(i,r),h_1,\dots,h_\delta$, where $h_j = H(i,v_j(r))$ for $1\leq j\leq \delta$. Add an arc $(i,r)\to (h_j, v_j(r))$ for each $1\leq j\leq \delta$. 

Since the messages are recovered correctly, the graph $\widetilde G$ will contain several disjoint copies of the expander graph $G$, say $G_{i_1},\dots,G_{i_k}$, though each $G_{i_j}$ is not necessarily aligned within the same column in $\widetilde G$. There will be arcs incoming to $G_{i_j}$ from nodes not in any $G_{i_j}$, but there are no outgoing arcs from $G_{i_j}$. In this case, we can recover each $G_{i_j}$ perfectly, and collect the full set $\{\mb{m}_{i_j,r}\}_{r=1}^{d_1}$ and thus recover $i_j$.
Let us rearrange the nodes within each row and align each copy of $G$ in the same column for clarity. In this case, the columns $i_1,\dots,i_k$ are exact copies of the expander graph $G$. See Figure~\ref{fig:ideal_recovery} for an illustration. 


The heavy hitters may not, however, be recovered in some repetitions and the messages could be seriously corrupted. When we are adding the arcs, we introduce two kinds of errors, respectively:
\begin{enumerate}\renewcommand{\labelenumi}{(\roman{enumi}) }
\setlength{\itemsep}{0pt}
\setlength{\parskip}{0pt}
\setlength{\parsep}{0pt} 
	\item We lose a node in $G_{i_j}$, i.e., the node is not present in $\widetilde G$ because the heavy hitter $i_j$ is not recovered in that repetition;
	\item We connect a node in $G_{i_j}$ to a node in some other $G_{i_{j'}}$ ($j\neq j'$), owing to errorous message.
\end{enumerate}
As before, we align each ``ideal copy'' of $G$ in the same column. See Figure~\ref{fig:real_recovery} for an example. We know that for a heavy hitter $i$, only a few messages $\{\mb{m}_{i,r}\}_r$ are ruined and the $i$-th column of $G_N$ will contain a large connected subgraph $G'$ of $G$, by Theorem~\ref{lem:graph_expander}. Hence, if we start a breadth-first search from an appropriate node with depth $c\log_\delta d_1$, the whole $G'$ will be visited. In other words, we shall obtain a large set of $\{p(i,r)\}$, only a small number of which will be associated with the same $i$, but we expect to obtain enough $\{p(i,r)\}$ of the same $i$, which turns out to be sufficient to extract $f$ associated with $i$ using a good error-correcting code such as the Parvaresh-Vardy code that allows us to recover the codeword from a large fraction of errors. Without attempting to identify the `appropriate node' described above, we shall perform this breadth-first search on every node in $\widetilde{G}$.
%


\paragraph{Guarantee} We shall show that the system described above meets the aforementioned guarantee. 
\begin{lemma}\label{lem:expander_recovery}
Let $\beta > 0$ and $\gamma\in(0,1)$ be constants. The encoding and decoding strategy of
Section~\ref{sec:expander_description} are correct in the sense of the
guarantee of that section, against the channel described in that
section. It uses $O(\epsilon^{-2} s\log N)$ measurements and runs in time $ O(s^{1+\beta}\poly(\log N,1/\epsilon))$, provided that $N=\Omega(\max\{s^2, s/\epsilon^2\})$ and $\epsilon = O\bigl((\frac{\log s}{\log N}\big)^\gamma\bigr)$. 
\end{lemma}
\begin{proof}
Combining Lemma~\ref{lem:two-layer}, Lemma~\ref{lem:one-layer} 
and Lemma~\ref{lem:two-layer-isolation}, one can show that there exists an $(4s,d_1d_2,\epsilon)$-bipartite expander such that
\begin{enumerate}\addtolength{\itemsep}{-0.25\baselineskip}
\renewcommand{\labelenumi}{(\alph{enumi})}
\setlength{\itemsep}{0pt}
\setlength{\parskip}{0pt}
\setlength{\parsep}{0pt}  	\item the bipartite expander exhibits a $(B_1,d_1,B_2,d_2)$ hashing structure, where the parameters are as in Lemma~\ref{lem:two-layer-isolation}, and each second-layer hashing satisfies $(O(s/\epsilon), O(\epsilon), O(1))$-isolation property;
	\item the bipartite expander satisfies the $(O(s/\epsilon), O(\epsilon), O(1))$-isolation property;
\end{enumerate}

As in the proof of Lemma~\ref{lem:decoy}, suppose that $|\mb{x}_i|\geq \epsilon/s$ for all $i\in\supp{\mb{y}}$, otherwise we can place the violated $i$'s into $\mb{z}$, causing $\|z\|_1$ to increase by at most $s\cdot \epsilon/s = \epsilon$, so we would have $\|z\|_1\leq 2$. Call the elements in $\supp{\mb{y}}$ heavy hitters. If $|\supp{\mb{y}}|\leq s/8$ our goal is automatically achieved, so we assume that $|\supp{\mb{y}}| > s/8$. 

\textbf{Step 1.} Overall we know from Lemma~\ref{lem:decoy} that we have at most $s/8$ decoys, or, we can recover $|\supp{\mb{y}}|-s/8$ heavy hitters from the second-layer bucket values, where successful recovery means that each of them dominates in at least $\alpha_2 d_1 d_2$ second-layer buckets, i.e., the bucket noise is at most $\nu=\epsilon/(2s)$.
For each of them, in at least $\beta_1 d_1$ of $d_1$ outer repetitions, it dominates in at least $\beta_2 d_2$ inner repetitions, where $(1-\beta_1)(1-\beta_2)>1-\alpha_2$. Because whenever an element dominates in the second-layer bucket, it must dominate the first-layer bucket incident to that second-layer bucket, we conclude that there exists a set $S\subseteq \supp{\mb{y}}$, $|S|\geq |\supp{\mb{y}}|-s/8$, such that each $i\in S$ dominates at least $\beta_1 d_1$ first-layer buckets among all $d_1$ repetitions, and in each of such repetitions, it dominates at least $\beta_2 d_2$ second-layer buckets.

We can choose the hidden constants in the bipartite expander parameters such that 
$\beta_1 \geq 1-\zeta$ and $\beta_2$ matches the error tolerance of the coding scheme we described in Lemma~\ref{lem:R-S_coding}, where $\zeta$ is the parameter we set in Section~\ref{sec:expander_description}.


\textbf{Step 2.} It follows from above that each $i\in S$ will be recovered in at least $\beta_1d_1$ outer repetitions, since its bucket value is $\geq \epsilon/s - \nu \geq \epsilon/(2s)$. Indeed, in every repetition of outer hashing, we collect top $O(s/\epsilon)$ (first-layer) buckets, so we will include every bucket with value $\geq \epsilon/(2s)$, and thus the heavy hitter $i$. In this case, the message associated with the heavy hitter will be recovered correctly, as the inner encoding can tolerate $1-\beta_2$ fraction of error. Therefore we know that for each $i\in S$, the associated messages will be correctly recovered in $\beta_1 d_1$ outer repetitions.

%

\textbf{Step 3.} As described in the previous section, we shall form a graph $\tilde G$. Note that for $i\in S$, $\beta_1 d_1$ nodes in the column are good nodes (i.e., with correct message). For each of them, perform a breadth-first search of $O(\log_\delta d_1)$ steps, collecting at most $d_1^c$ nodes. Since the column contains at most $(1-\beta)d_1 \leq \zeta d_1$ bad nodes, by Theorem~\ref{lem:graph_expander} and Property (d) of our choices of parameters, there exists a good node in the $i$-th column such that if we perform a breadth-first search of $c\log_\delta d_1$ steps, we shall collect $\alpha d_1$ good nodes which are all in the $i$-th column. The Parvaresh-Vardy code with our choice of parameters (Property (b) and (c)) enables us to include it in the list. We shall briefly describe the decoding below. Having collected at most $d_1^c$ points $(x,r(x))\in \F^{m+1}$, we consider all polynomials $Q(x,y_0,\dots,y_{m-1})$ of degree at most $d_X = \alpha d_1 - (h-1)m\log_{B_1}N$ in its first variable and at most $h-1$ in each such that $Q(x,r(x))=0$ for all $i$. Our choice of parameters (Property (c), i.e., $d_X h^m > d_1^c$) guarantees that such $Q$ exists. Then, the existence of $\alpha d_1$ good nodes (in the BFS visited nodes) indicates that the equation
\[
Q(x, f_i(x), (f_i^h \bmod{E})(x),\dots, (f_i^{h^{m-1}}\bmod{E})(x)) = 0
\]
has $\alpha d_1$ roots in $\F$ for $f_i$ corresponding to the coordinate $i\in S$. By our choice of parameters (Property (b)), the univariate polynomial $Q(x)$ has degree less than $\alpha d_1$ and must be identically zero. This means that $f_i(x)$ is a root of $Q^\ast(z) = Q(x,z,z^h,\dots,z^{h^{m-1}}) = 0$ over $\F[x]/E(x)$. We can find $f_i$ by factoring $Q^\ast$ and thus recover the position $i$ of the heavy hitter.

In the end, our candidate list will contain all $i\in S$, that is, we shall have recovered $|\supp{\mb{y}}|-s/8$ 
heavy hitters.

\paragraph{Number of Measurements} The number of measurements is $O(B_2d_1d_2)= O(\epsilon^{-2} s\log(N/s))$.

\paragraph{Size of Look-up Table} The inner decoding uses a look-up table of size $O(\log B_1) = O(\frac{s}{\epsilon} + \log\log N)$. The algorithm also stores the expander graph $G$, which takes space $O(d_1)$. Both are smaller than the space cost of the recovered graph $O(s d_1/\epsilon)$, so their contribution to the space complexity can be neglected.

\paragraph{Runtime} For each of $d_1$ repetitions, we shall recover every bucket with value $\geq \epsilon/(2s)$ in $O(B_1\log^3(B_1/k)) = O(s^{1+\beta}\poly(\log N,1/\epsilon))$ time. There are $O(s/\epsilon)$ of them in each repetition. Then we form a graph of size $O(sd_1/\epsilon)$. Forming this graph takes time $O(s^{1+\beta}\poly(\log N,1/\epsilon))$ from the argument above. Then we do breadth-first search of $c\log_\delta d_1$ steps on every node in $\widetilde{G}$. Each BFS takes $O(d_1^c)$ time. Each decoding of the BFS nodes takes $\poly(d_1,\log|B_1|)=\poly(\log N,1/\epsilon)$ time, and can be done deterministically (see, e.g., \cite[Theorem 4.3]{CEPR09}), since $|\F|$ has a small characteristic. Hence extracting heavy hitters $i$ from the recovered graph $\tilde G_N$ takes time $O(s\poly(\log N,1/\epsilon))$ and therefore, the overall runtime is $O(s^{1+\beta}\poly(\log N,1/\epsilon))$. In the end, we shall obtain a candidate list of size $O(s\poly(\log N,1/\epsilon))$. 
\end{proof}

\section{Toplevel System}\label{sec:toplevel}

Now we define a Toplevel system, similarly to~\cite{GLPS,PS12}, which is an algorithm that solves our overall problem. 
\begin{definition}
\label{def:toplevel}
An
{\em approximate sparse recovery system}
(briefly, a \emph{Toplevel} system),
consists of parameters $N$, $k$, $\epsilon$, an $m$-by-$N$
{\em measurement matrix} $\mtx{\Phi}$, and a
{\em decoding algorithm} $\mathcal{D}$ that satisfy the following property: for any vector $\signal\in \mathbb{R}^n$, given
$\mtx{\Phi}\signal$, the system
approximates $\signal$ by $\widehat \signal=\mathcal{D}(\mtx{\Phi}
\signal)$, which satisfies
\[
\|\widehat{\signal} - \signal\|_1
\le (1+\epsilon)\|\signal_{[k]} - \signal\|_1.
\]
\end{definition}

\begin{algorithm}[tb]
\caption{Toplevel System}
\label{alg:toplevel}
\begin{algorithmic}
\Require{$\mtx{\Phi}$, $\mtx{\Phi}\signal$, $N$, $k$, $\epsilon$}
\Ensure{$\widehat\signal$}
\State{$\widehat\signal\leftarrow0$}
\State{$\meas\leftarrow\mtx{\Phi}\signal$}
\For{ $j\gets 0$ to $\log k$}
  \State {Run Algorithm~\ref{algo:encoding_paradigm} on $\meas$ with length $N$, $s\leftarrow k/2^j$, $\eta\leftarrow \epsilon\alpha^j(1-\alpha)^2$ and obtain a candidate list $I$}
  \State {Run Algorithm~\ref{algo:weak} on candiate set $I$ with $s\leftarrow k/2^j$ and $\eta \leftarrow \epsilon \alpha^j(1-\alpha)$}
  \State {Let $\signal'$ be the result}
  \State {$\widehat\signal\gets \widehat\signal + \signal'$}
  \State {$\meas\gets \meas-\mtx{\Phi}\signal'$}
\EndFor
\State \Return $\widehat\signal$
\end{algorithmic}
\end{algorithm}

Using this definition, we restate our main result from Theorem~\ref{thm:mainresult} in a slightly different form.
\begin{theorem}
\label{thm:toplevel}
Let $\beta > 0$ and $\gamma \in (0,1)$ be constants. There exists $\alpha = \alpha(\beta,\gamma) \in (0, 1)$ such that Algorithm~\ref{alg:toplevel} yields a Toplevel system and 
uses $O(\epsilon^{-2}k\log N)$ measurements and runtime
$O(k^{1+\beta}\poly(\log N,1/\epsilon))$, provided that $N=\Omega(\max\{k^2, k/\epsilon^2\})$ and $\epsilon = O\bigl((\frac{\log k}{\log N})^\gamma\bigr)$. It also uses a look up table of size $O(\log N)$.
\end{theorem}
The proof follows easily using the results on the weak system. We need Lemma~\ref{lem:expander_recovery} for identification and Theorem~\ref{thm:weakI} for estimation. 

\begin{proof}
Suppose that in Lemma~\ref{lem:expander_recovery}, the exponent of $1/\epsilon$ in runtime is $c = c(\beta,\gamma) > 2$. Choose $\alpha < 1$ such that $\alpha^c > 1/2$. Assume that $\epsilon\leq 1/2$.

Using Lemma~\ref{lem:expander_recovery} for identification and Theorem~\ref{thm:weakI} for estimation, with appropriate choice of constants, we claim that at the beginning of the $j$-th step, $\mb{x} = \mb{y}+\mb{z}$, where $|\supp{\mb{y}}|\leq k/2^j$ and 
\[
\|\mb{z}\|_1\leq 1+\epsilon\left(1+\alpha+\alpha^2+\cdots+\alpha^{j-1}\right)(1-\alpha).
\]
We shall prove this claim by induction. Letting $s = k/2^j$, $\eta = \epsilon (1-\alpha)^2\alpha^j$ for identification, which introduces at most $\eta$ into the tail and the tail remains at most $3/2$ by assuming that all head items, i.e., the non-zero elements in $\mb{y}$, are all larger than $\eta/s$.

The identification procedure returns a candidate $I$ that contains $3/4$ fraction of $\supp{\mb{y}}$ (note that when the head is flat, we can change $\supp{\mb{y}}$ to be a superset that satisfies this condition without changing the norm of $\mb{z}$). Then the estimation procedure, with $s=O(k/2^j)$ and $\eta=\epsilon\alpha^{j+1}(1-\alpha)$ will give us 
\[
\mb{x} = \widehat{\mb{x}} + \widehat{\mb{y}} + \widehat{\mb{z}},
\]
where $|\supp{\mb{x}}|=O(s)$, $|\supp{\hat{\mb{y}}}|\leq s/2$ and 
\[
\|\widehat{\mb{z}}\|_1\leq \|\mb{z}\|_1 + \epsilon(1-\alpha)^2\alpha^j + \epsilon\alpha^{j+1}(1-\alpha) = \|\mb{z}\|_1 + \epsilon(1-\alpha)\alpha^j.
\]
It is easy to verify that $\|\hat{\mb{z}}\|_1\leq 1 + \epsilon \leq 3/2$ and thus Lemma~\ref{lem:expander_recovery} for identification and Theorem~\ref{thm:weakI} can be applied at the next round and the inductive hypothesis is satisfied. Therefore, in the end we obtain that
\[
\|\widehat{\mb{x}} - \mb{x}\|_1\leq \left(1+\epsilon\right)\|\mb{x}-\mb{x}_k\|_1.
\]
The number of measurements used for identification is
\[
O\left(\sum_j \frac{1}{\epsilon^2 \alpha^{2j}}\cdot \frac{k}{2^j}\log N\right) = O\left(\frac{k}{\epsilon^2}\log N\sum_j \left(\frac{1}{2\alpha^2}\right)^j\right) = O\left(\frac{k}{\epsilon^2}\log N\right)
\]
and the number of measurements used for estimation is
\[
O\left(\sum_j \frac{1}{\epsilon^2\alpha^j}\cdot \frac{k}{2^j}\log N\right) = O\left(\frac{k}{\epsilon^2}\log N\sum_j\left(\frac{1}{2\alpha}\right)^j\right) = O\left(\frac{k}{\epsilon^2}\log N\right)
\]
hence the total number of measurements is $O(\epsilon^{-2}k\log(N/k))$ as claimed.

It can be verified in a similar fashion that the total runtime is $O(k^{1+\beta} \poly(\log N,1/\epsilon))$, for which we need our choice of $\alpha$ satisfying that $\alpha^c > 1/2$.
\end{proof}

\begin{remark} We note that
\begin{enumerate}
\renewcommand{\labelenumi}{(\alph{enumi}) }
\setlength{\itemsep}{0pt}
\setlength{\parskip}{0pt}
\setlength{\parsep}{0pt} 
\item the constants in big $O$-notations and the power in $\poly(\log N,1/\epsilon)$ depend on $\beta$ and $\gamma$; 
\item as in Remark~\ref{rem:constraint_k}, The constraint that $k = O(\sqrt{N})$ could be weakened to $k = O(N^{1-\xi})$ for any $\xi > 0$;
\item the factor $k^{1+\beta}$ in the runtime is due to our choice of $B_1 = \Omega((k/\epsilon^2)^{1+\beta}\log(N/k))$ such that $\log B_1 = O(\log(B_1/k)) = O(d_2)$. When $k \leq (\log N)^c$ for some $c > 0$, since $B_1 = \Omega(k/\epsilon^{2(1+\beta)})$, choosing $B_1 = \Theta(k\log(N/k)/\epsilon^{2(1+\beta)})$ would suffice. It leads to runtime $O(k\poly(\log N,1/\epsilon))$.
\item For large $\epsilon$ we can take $d_1 = (\log N/\log(B_1/k))^{1+\alpha}$ for $\alpha > 0$, which gives an algorithm which uses more measurements $O(k\epsilon^{-2}\log^{1+\alpha} N)$ but suboptimal by only a logarithmic factor from the best known lower bound.
\end{enumerate}
\end{remark}

\section{Discussions and Open Problems}
\label{sec:closing}

\subsection{Codes}
At the core part of this paper lies the following list recovery problem: Suppose that there are $d_1 = \frac{1}{\epsilon}\cdot\frac{\log N}{\log(B/k)}$ lists $L_1,\dots, L_{d_1}$ with $|L_i| = O(k/\epsilon)$ for all $i=1,\dots,d_1$, we want to recover all possible codewords $c=(c_1,\dots,c_{d_1})$ such that $c_i\in L_i$ for at least $\Omega(d_1)$ different $i$s. It is natural to be tempted to apply Parvaresh-Vardy code directly without the expander structure. Indeed it works for some configurations of $k$ and $\epsilon$ with a runtime of $O(k\poly(\log N,1/\epsilon))$, but only for small $k$ and $\epsilon$. A direct application already fails even for $k=\exp(\sqrt{\log n})$. The runtime resulting from a direct application is also better for very small $k$, however, obtaining the precise range is difficult and beyond the scope of our work, as it relies on the precise complexity of factorizing a polynomial, which is not explicit in the literature.

Instead we use an expander structure to reduce the problem to $kd_1/\epsilon$ subproblems, each of which has a smaller number of nodes. Specifically the abstract problem is the following.

\begin{problem}
Let $\mathcal{C}$ be a $q$-ary code of block length $n$. For every sequence of subsets $S_1,\dots,S_n\subseteq [q]$ such that $\sum_{i=1}^n |S_i| \leq \ell$, find all codewords $(c_1,\dots,c_n)\in \mathcal{C}$ such that $|\{i: c_i\in S_i\}|\geq \alpha n$ in time $O(\poly(\ell, n,\log q))$.
\end{problem}

Note that instead of the usual assumption on individual size of each $S_i$ we have a bound on the sum of the sizes of $S_i$ here. Our choice of parameters is $q = \Theta(B_1)$, $n = d_1 = \Theta((\log_{B_1} N)/\epsilon)$, $\ell = \poly(d_1)$. The rate of the code is $\Theta(\epsilon)$. Our restrictions of $\epsilon$ comes from the application of the Parvaresh-Vardy code.  Potentially extractor codes~\cite{DBLP:journals/tit/Ta-ShmaZ04} and (in context) \cite{DBLP:journals/dam/Cheraghchi13} for context would yield improvement over this paper.

\subsection{Open Problems}
Below we list a few open problems.

\paragraph{Restriction on $\epsilon$.} The algorithm in this paper restricts $\epsilon$ to $(\frac{\log k}{\log N})^\gamma$ for any $\gamma > 0$ because of its way of applying the Parvaresh-Vardy code. In a sense our construction reduces the problem to a list recovery problem, as discussed in the previous subsection. We ask if it is possible to find an improvement by applying a better list recoverable code. 
The ultimate goal is to relax the restriction of $\epsilon$ to $\epsilon\leq \epsilon_0$ for some constant $\epsilon_0 > 0$.

\paragraph{Sparse Recovery in $\ell_2/\ell_1$ norm.} The ultimate problem is the $\ell_2/\ell_1$ problem with error guarantee as in \eqref{eqn:mixed-norm}.  We hope that the algorithm in this paper offers new ideas for the mixed-norm problem. Again the difficulty is in identification, as an RIP$_2$ matrix would be sufficient for estimation.


\paragraph{Post-measurement Noise.} In many algorithms on the sparse recovery problem, the input to the decoding algorithm is $\mtx{\Phi}\mb{x} + \nu$ instead of $\mtx{\Phi}\mb{x}$, where $\nu$ is an arbitrary noise vector. It is expected that our algorithm, with small changes if necessary, can tolerate substantial noise in $\ell_1$ norm, since the underlying structure is of a similar type to \cite{PS12}.
We leave to future work full analysis and possible improved algorithms.

\appendix
\section*{APPENDIX}
%
%

\section{Proof of Lemma~\ref{lem:one-layer}}

\begin{proof}
Let $p_s$ be the probability of a fixed set of $s$ elements hashed into less than $(1-\epsilon)ds$ elements. By symmetry this probability is independent of the $s$ positions and thus is well-defined. Hence the probability 
\begin{equation}\label{eqn:one-layer-a}
\Pr\{\text{hashing does not give an expander}\} = \sum_{s=2}^{4k} \binom{N}{s} p_s.
\end{equation}
Our goal is to show that
\begin{equation}\label{eqn:p_s}
p_s \leq \exp\left(-cs\ln\frac{eN}{s}\right)
\end{equation}
for some absolute constant $c>2$, for which it suffices to show that
\begin{equation}\label{eqn:one-layer-b}
p_s \leq \exp\left(-cs\ln\frac{N}{k}\ln\frac{Ck}{s}\right)
\end{equation}
for some $c, C > 0$. Indeed, it follows from \eqref{eqn:one-layer-b} that
\[
p_s \leq \exp\left(-cs\ln\frac{N}{k}\ln\frac{Ck}{s}\right) \leq \exp\left\{-cs\left(\ln\frac{N}{k}+\ln\frac{Ck}{s}\right)\right\} = \exp\left(-cs\ln\frac{CN}{s}\right)
\]
and \eqref{eqn:p_s} holds. Assume for the moment that \eqref{eqn:p_s} is proved, then we can bound \eqref{eqn:one-layer-a} to be
\begin{align*}
\sum_{s=2}^{\alpha k} \binom{N}{s} p_s 
&\leq \sum_{s=2}^{\alpha k} \exp\left\{s\ln\frac{eN}{s}-cs\ln\frac{CN}{s}\right\}\\
&\leq \sum_{s=2}^{\alpha k} \exp\left\{-(c-1)s\ln\frac{C'N}{s}\right\}\\
&\leq \sum_{s=2}^{\alpha k} \exp\left(-(c-1)s\log N\right) < \frac{1}{N^{c'}}
\end{align*}
as desired.

Now we compute $p_s$. Fix a set $S$ of $s$ elements. Suppose that they are hashed into $X_i$ ($i=1,\dots,d$) buckets in $d$ repetitions, respectively. We have that $1\leq X_i\leq s$ and $\sum X_i\leq (1-\epsilon)sd$. Define the event
\[
E_i(X_i) = \{S\text{ is hashed into }X_i\text{ rows in }i\text{-th reptition}\},
\]
and we shall compute $\Pr\{E_i(X_i)\}$.

When $E_i$ happens, there are $s-X_i$ repetitions. Consider we hash the element one by one, choosing $b_1,\dots,b_d\in \{1,\dots,B\}$ sequentially. We have a collision when selecting $b_i$ if $b_i\in \{b_1,\dots,b_{i-1}\}$. The probability that a collision occurs at step $i$, even conditioned on $b_1,\dots,b_{i-1}$, is at most $i/B\leq s/B$. Therefore,
\[
\Pr\{E_i(X_i)\}\leq \binom{s}{s-X_i}\left(\frac{s}{B}\right)^{s-X_i}.
\]
Hence
\[
p_s = \sum \Pr\{E_1(X_1),\dots,E_d(X_d)\} = \sum \prod_{i=1}^d \binom{s}{s-X_i}\left(\frac{s}{B}\right)^{s-X_i} = \sum \left(\frac{s}{B}\right)^{sd-\sum X_i} \prod_{i=1}^d \binom{s}{s-X_i}
\]
where the summation is over all possible configurations of $\{X_i\}$. Invoking the combinatorial identity
\begin{equation}\label{eqn:comb_identity}
\sum_{k_1+k_2+\cdots+k_m=n} \binom{r_1}{k_1}\binom{r_2}{k_2}\cdots\binom{r_m}{k_m} = \binom{r_1+r_2+\cdots+r_m}{n}
\end{equation}
and writing $X=\sum X_i$, we see that
\[
p_s \leq \sum_{X=d}^{(1-\epsilon)sd} \left(\frac{s}{B}\right)^{sd-\sum X_i}\binom{sd}{sd-\sum X_i} \leq \sum_{X=\epsilon sd}^d \binom{sd}{X}\left(\frac{s}{B}\right)^{X}
\]
Now we invoke Chernoff bound
\begin{equation}\label{eqn:chernoff_2}
\sum_{k=\epsilon n}^n \binom{n}{k}\lambda^k \leq \left(\frac{e\lambda}{\epsilon}\right)^{\epsilon n},\quad \lambda<\epsilon
\end{equation}
to obtain that
\[
p_s\leq \left(\frac{es}{\epsilon B}\right)^{\epsilon sd}\leq \exp\left(-cs\log\frac{N}{k}\ln\frac{Ck}{s}\right)
\]
as desired, where the constants $c,C > 0$ can be made arbitrarily big.
\end{proof}

\section{Proof of Lemma~\ref{lem:one-layer-isolation}}

\begin{proof}
Let $S$ be a set of size $s\leq L$. We shall bound the probability $p_s$ (which is defined by symmetry) that at least $\epsilon s$ elements of $S$ collide with each other in at least $\zeta d$ repetitions. When this happens, there are at least $\epsilon\zeta ds$ colliding element-repetition pairs. As in Lemma~\ref{lem:one-layer} it suffices to have \eqref{eqn:one-layer-b} for some $c,C > 0$ that can be made arbitrarily large.

In one repetition, one element of $S$ collide with others with probability $\leq s/B$. By a coupling argument as in \cite{PS12}, among all $sd$ element-repetition pairs with expected $\mu = s^2 d/B$ failed pairs, there are at least $\zeta \epsilon sd$ failed pairs with probability
\[
\left(\frac{e\mu}{\zeta\epsilon ds}\right)^{\zeta\epsilon sd} = \left(\frac{es}{\zeta\epsilon B}\right)^{\zeta\epsilon sd}\leq \exp\left(-cs\log\frac{N}{k}\ln\frac{Ck}{s}\right)
\]
as desired, where the absolute constants $C,c>0$ can be made arbitrary large. 
\end{proof}

\section{Proof of Lemma~\ref{lem:two-layer}}

\begin{proof}
Let $p_s$ be the probability of a fixed set of $s$ elements hashed into less than $(1-\epsilon)ds$ elements. By symmetry this probability is independent of the $s$ positions and thus is well-defined. Hence the probability 
\begin{equation}\label{eqn:7a}
\Pr\{\text{hashing does not give an expander}\} = \sum_{s=2}^{4k} \binom{N}{s} p_s.
\end{equation}
Similarly to Lemma~\ref{lem:one-layer}, it suffices to show that
\begin{equation}\label{eqn:7b}
p_s \leq \exp\left(-cs\ln\frac{N}{k}\right)
\end{equation}
Assume for the moment that this is proved, then we can bound \eqref{eqn:7a} to be
\begin{align*}
\sum_{s=2}^{4k} \binom{N}{s} p_s 
&\leq \sum_{s=2}^{4k} \exp\left\{s\ln\frac{eN}{s}-cs\ln\frac{N}{k}\right\}\\
&\leq \sum_{s=2}^{4k} \exp\left\{s\ln(eN)-\frac{c}{2}s\ln(eN)\right\}\quad (k\leq \sqrt{N/e})\\
&\leq \sum_{s=2}^{4k} \exp\left(-\left(\frac{c}{2}-1\right)s\log(eN)\right) < \frac{1}{N^{c'}}
\end{align*}
as desired.

Now we prove \eqref{eqn:7b}. Fix a set $S$ of $s$ elements. The outer layer of hashing has $d_1$ blocks of size $B_1$, and let $Y_i$ ($i=1,\dots,d_1$) be the number of hashed row of the $s$ elements in $i$-th block. The inner layer has $d_1d_2$ blocks, indexed by $(i,j)_{1\leq i\leq d_1,1\leq j\leq d_2}$ of size $B_2$, and let $X_{ij}$ be the number of hashed row of the $s$ elements in the $(i,j)$-th block. Define the events
\begin{gather*}
E_i(Y_i) = \{S\text{ is hashed into }Y_i\text{ rows in }i\text{-th outer block}\}\\
E_{ij}(X_{ij}) = \{S\text{ hashed into }X_{ij}\text{ rows in }(i,j)\text{-th inner block}\}
\end{gather*}
First we calculate $\Pr\{E_i\}(Y_i)$. Consider we pick a row at one time for an element in $S$ in order. When $E_i(Y_i)$ happens there are at least $s-Y_i$ collisions, hence
\[
\Pr\{E_i(Y_i)\} \leq \binom{s}{s-Y_i} \left(\frac{s}{B_1}\right)^{s-Y_i} 
\]
and similarly
\[
\Pr\{E_{ij}(X_{ij})|E_i(Y_i)\} \leq \binom{Y_i}{Y_i-X_{ij}} \left(\frac{Y_i}{B_2}\right)^{Y_i-X_{ij}} 
\]
It follows that
\begin{align*}
p_s
&=   \sum  
     \Pr\{E_{11}(X_{11}),\dots,E_{d_1d_2}(X_{d_1d_2})|E_1(Y_1),\dots,E_{d_1}(Y_{d_1})\}\Pr\{E_1(Y_1),\dots,E_{d_1}(Y_{d_1})\}\\  
&\leq 
     \sum \prod_i \Pr\{E_i\}(Y_i) \prod_{i,j} \Pr\{E_{ij}(X_{ij})|E_i(Y_i)\} \\
& \leq \sum \prod_i \binom{s}{Y_i} \left(\frac{s}{B_1}\right)^{s-Y_i} \cdot
            \prod_{i,j} \binom{Y_i}{X_{ij}}\left(\frac{Y_i}{B_2}\right)^{Y_i-X_{ij}}\\
& \leq \sum \left(\frac{s}{B_1}\right)^{sd_1-\sum Y_i} \left(\frac{s}{B_2}\right)^{d_2\sum Y_i-\sum X_{ij}} \prod_i \binom{s}{Y_i}  \prod_{i,j} \binom{Y_i}{X_{ij}}
\end{align*}
where the summation is taken over all possible configurations of $\{X_i\}$ and $\{Y_i\}$ so that $s\geq Y_i\geq \max_j X_{ij}$ and $\sum X_{ij} \leq (1-\epsilon)sd_1d_2$.

Invoking the combinatorial equality \eqref{eqn:comb_identity}
and letting $X=\sum X_{ij}$ and $Y=\sum Y_i$, we obtain that
\begin{align}
p_s &\leq \sum_{Y=d_1}^{sd_1}\binom{sd_1}{Y} \left(\frac{s}{B_1}\right)^{sd_1-Y} \sum_{X=d_1d_2}^{\min\{d_2Y,(1-\epsilon)sd_1d_2\}} \binom{d_2Y}{X}\left(\frac{s}{B_2}\right)^{d_2Y-X}\notag\\
&\leq \sum_{Y=d_1}^{(1-\epsilon/2)sd_1}\binom{sd_1}{Y}\left(\frac{s}{B_1}\right)^{sd_1-Y} \sum_{X=d_1d_2}^{d_2Y} \binom{d_2Y}{X}\left(\frac{s}{B_2}\right)^{d_2Y-X}\notag\\
&\qquad  + \sum_{Y=(1-\epsilon/2)sd_1}^{sd_1}\binom{sd_1}{Y}\left(\frac{s}{B_1}\right)^{sd_1-Y} \sum_{X=d_1d_2}^{(1-\epsilon)sd_1d_2} \binom{d_2Y}{X}\left(\frac{s}{B_2}\right)^{d_2Y-X}\notag\\
&=: S_1+S_2\label{eqn:7c}
\end{align}
We bound $S_1$ and $S_2$ separately. First,
\[
S_1 \leq \sum_{Y=d_1}^{(1-\epsilon/2)sd_1}\binom{sd_1}{Y}\left(\frac{s}{B_1}\right)^{sd_1-Y}\left(1+\frac{s}{B_2}\right)^{d_2Y}\\
\leq \left(1+\frac{s}{B_2}\right)^{sd_1d_2}\sum_{Y=\frac{\epsilon}{2} sd_1}^{sd_1}\binom{sd_1}{Y}\left(\frac{s}{B_1}\right)^{Y} 
\]
It follows from Chernoff bound \eqref{eqn:chernoff_2} that
\begin{align}
S_1 &\leq \left(1+\frac{s}{B_2}\right)^{sd_1d_2} \left(\frac{es}{\frac{\epsilon}{2} B_1 }\right)^{\epsilon sd_1/2}\notag\\
&\leq \exp\left\{-\frac{1}{2}\epsilon sd_1\left(\ln\frac{\epsilon B_1}{2es}\right) + sd_1d_2\ln\left(1+\frac{s}{B_2}\right)\right\}\notag\\
&\leq \exp\left\{-\frac{1}{4}\epsilon sd_1\ln\frac{B_1}{k}+c_2\epsilon sd_1d_2\right\} \quad (\text{since }B_1 \gtrsim k/\epsilon^2)\notag\\
&\leq \exp\left\{-c_3 s\ln\frac{N}{k}\right\}\label{eqn:7s1}
\end{align}
where the absolute constant $c_2 > 0$ can be made arbitrarily close to $0$ and the absolute constant $c_3$ can be made arbitrarily large.

Now we bound $S_2$. When $Y\geq (1-\epsilon/2)sd_1$ then
\[
\frac{(1-\epsilon)sd_1d_2}{d_2Y}\leq 1-\frac{\epsilon}{2}.
\]
Again invoking Chernoff bound,
\[
\sum_{X=d_1d_2}^{(1-\epsilon)sd_1d_2} \binom{d_2Y}{X}\left(\frac{s}{B_2}\right)^{d_2Y-X} \leq \left(\frac{es}{\frac{\epsilon}{2} B_2}\right)^{d_2Y - (1-\epsilon) sd_1d_2} \leq \left(\frac{s}{C'k}\right)^{d_2Y - (1-\epsilon)sd_1d_2}
\]
where $C'>0$ is an absolute constant which can be made arbitrarily large. So
\begin{align*}
S_2 &\leq \sum_{Y=(1-\epsilon/2)sd_1}^{sd_1}\binom{sd_1}{Y}\left(\frac{s}{B_1}\right)^{sd_1-Y} \left(\frac{s}{C'k}\right)^{\epsilon sd_1d_2/2}\\
&\leq \sum_{Y=0}^{(\epsilon/2)sd_1}\binom{sd_1}{Y}\left(\frac{s}{B_1}\right)^{Y} \left(\frac{s}{C'k}\right)^{\epsilon sd_1d_2/2}\\
&\leq 2 \left(\frac{s}{C'k}\right)^{\epsilon sd_1d_2/2}
\end{align*}
It immediately follows, similarly to upper-bounding $S_1$, that
\begin{equation}\label{eqn:7s2}
S_2 \leq \exp\left\{-c_4s\ln\frac{N}{k}\ln\frac{C'k}{s}\right\},
\end{equation}
where $c_4 > 0$ can be made arbitrarily large. Plugging \eqref{eqn:7s1} and \eqref{eqn:7s2} into \eqref{eqn:7c} we see that \eqref{eqn:7b} holds. This completes the proof.
\end{proof}

\section{Proof of Lemma~\ref{lem:two-layer-isolation}}
\begin{proof}
Fix a set $S$ of size $s$. Let event $\mathcal{E}$ be that at least $(1-\epsilon/2)s$ elements in $S$ are isolated in at least $(1-\zeta/2)d_1$ first-layer buckets. Similarly to Lemma~\ref{lem:one-layer-isolation} we know that
\[
\Pr\{\mathcal{E}^c\} \leq 
\left(\frac{c's}{\zeta\epsilon B_1}\right)^{\zeta\epsilon sd_1}\leq e^{-cs\log\frac{N}{k}}
\]
where $c'$ is an absolute constant and $c>0$ can be made arbitrarily large. In the above we used that fact that since $B_1=\Omega(k/(\zeta^{\alpha}\epsilon^{2\alpha}))$ it holds that
\[
\ln\frac{\zeta\epsilon^2 B_1}{c_1 k} \geq \left(1-\frac{1}{\alpha}\right)\ln\frac{B_1}{k}.
\]

We condition on event $\mathcal{E}$. Among the $(1-\epsilon/2)s$ elements we shall show that at least $(1-\epsilon)$ of them are isolated in at least $(1-\zeta)d_1d_2$ second-layer buckets. That means, there are a total of at least $\frac{\epsilon}{2}\frac{\eta}{2} sd_1d_2$ failed element-reptitions. But now, the probability of each collision is always bounded by $s/B_2$ even conditioned on previous outcomes, and we can proceed as in Lemma~\ref{lem:one-layer-isolation} to conclude that there are at least $\theta\zeta\epsilon sd_1d_2$ (for some absolute constant $\theta$) with probability at most
\[
\left(\frac{es}{\theta\zeta\epsilon B_2}\right)^{\theta \zeta\epsilon sd_1d_2} \leq e^{-c''s\log\frac{N}{k}},
\]
as desired, where the constant $c'' > 0$ can be made arbitrarily large.
\end{proof}

\bibliographystyle{ACM-Reference-Format-Journals}
\bibliography{literature}

\begin{acks}
We thank the anonymous reviewer for the valuable comments and suggestions that greatly contributed to improving this paper.
\end{acks}
\end{document}